\documentclass[]{pasj01}

\begin{document} 
\Received{2017/04/08}
\Accepted{2017/08/11}
\title{Enhancement of Galaxy Overdensity around Quasar Pairs at $z<3.6$ based on the Hyper Suprime-Cam Subaru Strategic Program Survey}

\author{Masafusa \textsc{Onoue}\altaffilmark{1,2}, 
Nobunari \textsc{Kashikawa}\altaffilmark{1,2},
Hisakazu \textsc{Uchiyama}\altaffilmark{1,2},
Masayuki \textsc{Akiyama}\altaffilmark{3},
Yuichi \textsc{Harikane}\altaffilmark{4,5},
Masatoshi \textsc{Imanishi}\altaffilmark{1,2},
Yutaka \textsc{Komiyama}\altaffilmark{1,2}, 
Yoshiki \textsc{Matsuoka}\altaffilmark{1,2,6},
Tohru \textsc{Nagao}\altaffilmark{6},
Atsushi J. \textsc{Nishizawa}\altaffilmark{7},
Masamune \textsc{Oguri}\altaffilmark{8,9,5},
Masami \textsc{Ouchi}\altaffilmark{4,5},
Masayuki  \textsc{Tanaka}\altaffilmark{1,2},
Yoshiki \textsc{Toba}\altaffilmark{10},
Jun \textsc{Toshikawa}\altaffilmark{2}
}%
\altaffiltext{1}{Department of Astronomical Science, Graduate University for Advanced Studies (SOKENDAI), 2-21-1, Osawa, Mitaka, Tokyo 181-8588}
\altaffiltext{2}{National Astronomical Observatory of Japan, 2-21-1, Osawa, Mitaka, Tokyo 181-8588}
\altaffiltext{3}{Astronomical Institute, Tohoku University, Aramaki, Aoba, Sendai, Miyagi 980-8578}
\altaffiltext{4}{Institute for Cosmic Ray Research, The University of Tokyo, Kashiwa-no-ha, Kashiwa 277-8582}
\altaffiltext{5}{Kavli Institute for the Physica and Mathematics of the Universe (Kavli IPMU), WPI, The University of Tokyo, Kashiwa, Chiba 277-8583}
\altaffiltext{6}{Research Center for Space and Cosmic Evolution, Ehime University, Matsuyama, Ehime 790-8577}
\altaffiltext{7}{Institute for Advanced Research, Nagoya University, Furo-cho, Chikusa-ku, Nagoya 464-8602}
\altaffiltext{8}{Research Center for the Early Universe, University of Tokyo, Tokyo 113-0033}
\altaffiltext{9}{Department of Physics, University of Tokyo, Tokyo 113-0033}
\altaffiltext{10}{Academia Sinica Institute of Astronomy and Astrophysics, P.O. Box 23-141, Taipei 10617, Taiwan}
\email{masafusa.onoue@nao.ac.jp}


\KeyWords{quasars: galaxies: general } 

\maketitle

\begin{abstract}
We investigate the galaxy overdensity around proto-cluster scale quasar pairs
at high ($z>3$) and low ($z\sim1$) redshift
based on the unprecedentedly wide and deep optical survey of the Hyper Suprime-Cam Subaru Strategic Program (HSC-SSP).
Using the first-year survey data covering effectively $\sim121$ deg$^2$ with the $5\sigma$ depth of $i\sim26.4$ and the SDSS DR12Q catalog,
we find two luminous pairs at $z\sim3.3$ and $3.6$ which reside in $>5\sigma$ overdense regions of $g$-dropout galaxies at $i<25$.
The projected separations of the two pairs are $R_\perp=1.75$ and $1.04$  proper Mpc, and their velocity offsets are $\Delta V=692$ and $1448$ km s$^{-1}$, respectively.
This result is in clear contrast to the average $z\sim4$ quasar environments as discussed in \citet{Uchiyama17} 
and implies that the quasar activities of the pair members are triggered via major mergers in proto-clusters, 
unlike the vast majority of isolated quasars
in general fields that may turn on via non-merger events such as bar and disk instabilities.
At $z\sim1$, we find $37$ pairs with $R_\perp<2$ pMpc and $\Delta V<2300$ km s$^{-1}$ in the current HSC-Wide coverage, including four from \citet{Hennawi06}.
The distribution of the peak overdensity significance within two arcminutes around the pairs has a long tail toward high density ($>4\sigma$) regions.
Thanks to the large sample size, we find a statistical evidence that this excess is unique to the pair environments when compared to single quasar and randomly selected galaxy environments at the same redshift range.
Moreover, there are nine small-scale ($R_\perp<1$ pMpc) pairs, two of which are found to reside in cluster fields.
Our results demonstrate that  $<2$ pMpc-scale quasar pairs at both redshift range
tend to occur in massive haloes, 
although perhaps not the most massive ones,
and that they are useful to search for rare density peaks.

\end{abstract}

\section{Introduction}
The subject of how active galactic nuclei (AGN) connect to their surrounding galaxy formation and host dark matter haloes has been a matter of debate in modern extragalactic astronomy.
When one assumes that supermassive black holes (SMBHs) grow primarily via gas accretion,
the triggering mechanism of luminous quasars would be major mergers of gas-rich galaxies, which transform plenty of cold gas into the central SMBHs (e.g., \cite{Hopkins08, AlexHic12}).
The $M_\mathrm{BH}-\sigma_*$ relation  (e.g., \cite{Magorrian98, KorHo13}) suggests a co-evolutionary growth history of SMBHs and host galaxies throughout the cosmic epoch in which massive galaxies host large SMBHs.

To investigate how the most mature clusters in the local universe are formed along with the hierarchal large-scale structure formation, 
AGN have been used for proto-cluster searches at high redshift.
This is motivated by the assumption that galaxies in high density regions have earlier episodes of star formation and SMBH growth than those in normal and poor environments.
Since gas-rich major mergers would more frequently happen in massive haloes than in less massive ones, luminous AGN can be used as landmarks of proto-clusters, while the blind searches of such rare density peaks requires a wide-field observation.
There are various studies testing the validity of using AGN for proto-cluster searches.
For example,\citet{Venemans07} clearly show that radio galaxies are likely to be associated with proto-clusters at $z>2$ based on their Ly$\alpha$ emitter (LAE) searches.
\citet{Wylezalek13} show that radio-loud AGN prefer massive environments using sources selected by the Infrared Array Camera \citep{Fazio04} of the {\it Spitzer Space Telescope}, suggesting that the high density environments induce high spins of the SMBHs and radio jets enhancement.
On the other hand, environment studies on the highest-redshift quasars show that luminous $z>6$ quasars are not necessarily associated with rich environments \citep[and reference therein]{Mazzucchelli17}.
Moreover, a deep and large spectroscopic sample of $2<z<3$ quasars from the Baryon Oscillation Spectroscopic Survey (BOSS, \cite{Dawson13})  imply that the redshift evolution of the quasar 
auto-correlation signal gets flattened from lower redshift \citep{Eftekharzadeh15}.
Although their result is inconsistent with a $z>3$ study by \citet{Shen07} and the reason is unclear,
it may imply that highly energetic feedback from quiescent AGN suppresses further cold-gas assembly onto the SMBHs in the most massive haloes (``radio-mode" feedback).
Thereby, these arguments follow that the host halo of a luminous quasar at high redshift is not the most massive.
This scenario is supported by several semi-analytical studies (e.g., \cite{Fanidakis13, Orsi15}).

The Hyper Suprime-Cam (HSC, \cite{Miyazaki12}) is a new optical camera installed at the prime focus of the Subaru telescope.
With the $8.2$m mirror of the Subaru, ten minutes imaging goes as deep as $r_\mathrm{lim,5\sigma}\sim26$, which is three magnitudes deeper than the Sloan Digital Sky Survey (SDSS, \cite{York00}).
The most characteristic feature of the HSC is its gigantic field-of-view of $1.5$ degree diameter.
Taking advantage of the survey efficiency of the camera, the HSC Subaru Strategic Program (HSC-SSP) covers $1400$ deg$^2$ with its five broad-band filters in the Wide layer\footnote{For more information of the HSC-SSP survey, we encourage readers to refer to the public data release paper \citep{DRpaper} and the survey design paper \citep{SSPpaper}. 
The filter information is given in \citet{Filter}}.
This survey is a five-year survey, which began in March 2014.
Its first $\sim100$ deg$^2$ data of the Wide, Deep and UltraDeep layers has been open to public since Febrary 2017 \citep{DRpaper}.
As the survey has started the enormously wide-field observation, \citet{Toshikawa17} have found 179 promising proto-cluster candidates at $z\sim3.8$ over $121$ deg$^2$ currently available among the collaboration.
This number overwhelms that of the previously known proto-clusters at the same redshift range, enabling them to investigate the clustering of proto-clusters for the first time.
Their $g$-dropout catalog which efficiently selects Lyman break galaxies is used for a quasar environment study by \citet{Uchiyama17}, which is our companion paper. 
They exploit the catalog to examine how well quasars at $3.3<z<4.2$ trace the HSC proto-clusters based on their $151$ BOSS quasar sample,
finding that the luminous quasars in general do not reside in rich environments when compared to $g$-dropout galaxies.
In particular, there are only six cases where the individual luminous quasars are hosted in the HSC proto-clusters within three arcminiutes ($\sim1.3$ proper Mpc) from the density peaks.
Their result sheds a new light on the triggering mechanism of luminous quasars from an environmental point of view; the quasar activity at high redshift is more common in general environments than previously thought. 
One interpretation would be that the major merger is not a unique mechanism to trigger a luminous quasar and 
other channels, such as the secular process (bar and disk instabilities), which closes in its own system and has less to do with its environment, play a significant role.
In fact, recent studies have shown that the morphology of luminous-quasar hosts at $z<2$ are not highly distorted and not significantly different from that of inactive galaxies (e.g., \cite{Mechtley16, Villforth17}).
Another way to explain their result is the strong AGN feedback, which suppresses star formation in the vicinity of quasars.
It is also possible that the majority of star-forming galaxies around $z\sim 4$ quasars are dusty and thus highly obscured, for which optical-based selection completeness is low.

When one assumes that major mergers dominate the role of triggering quasars in massive haloes,
it is likely that environments around multiple quasars, i.e., a physical association of more than one quasar in close separation are more biased and they are much more efficient in pinpointing proto-clusters.
While such populations are extremely rare, there are several studies on the pair and multiple quasars.
\citet{Djorgovski87} first report the discovery of a binary quasar at $z=1.345$, a radio source PKS 1145-071, separated by $4.2$ arcseconds.
A small number of $z>4$ quasar pairs with less than $1$ pMpc projected separation has been reported \citep{Schneider00, Djorgovski03, McGreer15}.
Based on the SDSS, \citet{Hennawi06} and \citet{Hennawi10} construct large catalogs of spectroscopically confirmed quasar pairs at $z<3$ and $3<z<4$, respectively.
Quasar pairs with sub pMpc-scale separation like their samples are of particular interest for the study of the small-scale clustering of quasars.
Several papers argue that quasar clustering is enhanced in such a small scale, which is suggestive of enhanced quasar activity in rich environments (e.g., \cite{Hennawi06, Eftekharzadeh17}, but also see \cite{Kayo12}).
Regarding the environments, most of the previous studies focus on $z<1$ pairs with $\sim10$ sample sizes  (e.g., \cite{Boris07, Farina11, Green11, Sandrinelli14}).
While the observation depth and the target selection differ among these studies, 
they all suggest that the extremely rare quasar pairs are not always associated with significant overdensity of galaxies.
At high-redshift, \citet{Fukugita04} look for the overdensity enhancement around a $z=4.25$ quasar pair, SDSS J1439-0034, but see no significant difference from a general field within a $5.8$ arcminutes$^2$ area.
On the other hand, it is remarkable that \citet{Hennawi15} show an exceptionally strong enhancement of the LAE surface density around a $z=2$ quasar quartet (``Jackpot nebulae"), with more than an order of excess on $<100$ pkpc scale, demonstrating the extremeness of multiple quasar environments.
Moreover, \citet{Cai17} find an extremely massive overdensity of LAEs at $z=2.32$ which is associated with multiple quasars.

This paper focuses on the galaxy overdensity around pairs of the BOSS quasars up to $z\sim3.6$, thanks to the unprecedented coverage and depth of the HSC-SSP dataset.
The outline of this paper is as follows:
In Section~\ref{sec:data_highz}, we describe our quasar pair sample at high redshift ($3.3<z<4.2$) and introduce how we select surrounding galaxies using the HSC-SSP photometric dataset.
In Section~\ref{sec:results_z34}, we show that two quasar pairs at $z\sim3.3$ and $\sim3.6$ are both associated with the HSC proto-clusters,
supporting that the rare occurrence of $<2$ pMpc-scale pairs traces rich environments.
Section~\ref{sec:lowz_pair} describes our overdensity measurements around $37$ quasar pairs at $z<1.5$ including four pairs from literature, using the photometric redshift catalog of the HSC-SSP.
We find statistical evidence that a significantly higher fraction of quasar pairs resides in dense environments than single quasars and randomly selected galaxies, which is described in Section~\ref{sec:lowz_pair_res}.
We discuss the quasar pair environments in Section~\ref{sec:discussion}, especially comparing with the study of single quasar environments at $z\sim3.8$ \citep{Uchiyama17}.
Finally, the summary is given in Section~\ref{sec:summary}.

Throughout this paper, we adopt a $\Lambda$CDM cosmology with $H_0=70$ km s$^{-1}$ Mpc$^{-1}$, $\Omega_m=0.3$ and $\Omega_\Lambda=0.7$.
Unless otherwise stated, the magnitudes cited in this paper are the CModel magnitude (in AB system), which is derived by fitting images with combination of the exponential and de Vaucouleurs profiles. 
The CModel magnitude is conceptionally the same as the PSF magnitude for point sources.

\section{Data and Sample Selection}\label{sec:data_highz}
\subsection{Effective Region in the HSC-Wide Dataset}\label{sec:eff_region}
In this paper, we use a photometric catalog of the HSC-SSP Wide survey for our surrounding galaxy selection.
The photometric catalog using a dedicated pipeline (hscPipe; \cite{pipeline}) has been opened to the collaboration,
the latest dataset of which is denoted as DR S16A with its Wide component covering $\sim170$ deg$^2$ in five broad-band filters ($grizy$).
This area consists of several large fields along the equator (W-GAMA09H, W-Wide12H, W-GAMA15H, W-VVDS, W-XMMLSS) and one field at Decl.$=43$ deg (W-HECTMAP).
The average 5$\sigma$-limiting magnitudes\footnote{this limiting magnitude is defined as the point where the PSF magnitude has $S/N\sim5\sigma$ at $0.5-0.7$ arcsec seeing condition. See \citet{DRpaper}} are as follows:
$g\sim26.8$, $r\sim26.4$, $i\sim26.4$, $z\sim25.5$, and $y\sim24.7$.
We share the $g$-dropout catalog of the HSC proto-cluster search project \citep{Toshikawa17}.
To take into account the partly inhomogeneous imaging in each HSC filter over the large survey regions, they carefully remove area with shallow depths in either $g$-, $r$-, or $i$-bands based on sky noise measurements at each $12\times12$ arcmin$^2$ sub-regions ({\it ``patch"} in the HSC-SSP term), as well as the masked regions, for example, around saturated stars, at the edge of the images, and on bad pixels based on the photometry flags of the hscPipe (see Section~2 of \cite{Toshikawa17} for more details).
As a result, they construct a highly clean and uniform $g$-dropout galaxy sample over a total effective area of $\sim121$ deg$^2$.
Specifically, W-GAMA09H field is discarded in our analysis, because the number counts of the $g$-dropout galaxies has an offset compared to the other fields at bright magnitude range due to its shallow depth in $r$-band.

\subsection{Quasar pair sample at $3<z<4$}
In this section, we present our selection and sample of quasar pairs at $3<z<4$.
We use the latest catalog of spectroscopically confirmed quasars from the SDSS-{\sc III} BOSS survey (DR12Q, \cite{Paris17}),
which contains about $300$ thousands quasars in $9376$ deg$^2$ down to $g=22.0$ or $r=21.85$.
While the BOSS survey originally targets quasars at $2.15\leq z \leq 3.5$, the redshift distribution of the DR12Q sample has a wide skirt up to $z\sim6.4$.
Moreover, the DR12Q catalog has a secondary redshift peak at $z\sim0.8$, which is due to the SDSS-color similarity of quasars at this redshift range to that of $2<z<3$ quasars, enabling us to also investigate low-redshift quasar environments as described in Section \ref{sec:lowz_pair}.
We require secure redshift determination using a flag given in the DR12Q catalog (i.e., {\tt ZWARNING=0}).

Since we assume such a situation in which more than one quasar emerges in the same massive halo,
we define quasar pairs as two quasars with their separation closer than the size of massive proto-clusters.
We extract quasar pairs from the DR12Q catalog following a framework given by a simulation done by \citet{Chiang13}.
According to their definition of the most massive proto-clusters, which are the progenitors of $M_\mathrm{halo, z=0}>10^{15} M_\odot$ clusters,
and their characteristic size at the concerned redshift,
we extract quasar pairs with their projected proper distance of $R_\perp<4$ pMpc and velocity offset of $\Delta V<3000$ km s$^{-1}$.
This definition is more relaxed than previous studies such as \citet{Hennawi10}, since
they assume that the pairs are gravitationally bound systems with $R_\perp<1$ pMpc.
The redshift range is limited to $3.3<z<4.2$ where the selection completeness of the HSC $g$-dropouts is over $0.4$ \citep{Ono17}.
The Lyman break of galaxies at this range is shifted to $r$-band, and therefore $g-r$ and $r-i$ colors can be used to distinguish those galaxies from contaminants such as main-sequence stars and $z<1$ galaxies. 
The velocity offset between the quasar members in a pair $\Delta V$ is determined from the SDSS's visually inspected redshifts ({\tt Z\_VI}).
Considering the uncertainty of the BOSS redshifts primarily relying on the Lyman break and Ly$\alpha$ emission ($\sim1000$ km s$^{-1}$), possible peculiar motion between the pair members ($\sim500$ km s$^{-1}$),
and also their physical separation in the radial direction, we apply $\Delta V_\mathrm{max}=3000$ km s$^{-1}$ as the maximum velocity offset of the $z\sim3.8$ pairs.
Finally, the BOSS quasar spectra are visually checked to confirm their secure classification and redshift determination.

As a result, two pairs of quasars are found at $z\sim3.6$ and $z\sim3.3$ from the DR12Q catalog.
Table \ref{tab:QSOP_z34} lists the two pairs with their angular separation $\Delta \theta$, projected separation $R_\perp$ and velocity offset $\Delta V$, in which the average redshift of the two quasars in pairs is regarded as a pair redshift.
QSOP1 is a pair of quasars at $z=3.585$ and $z=3.574$ with their angular separation $\Delta\theta=241$ arcseconds ($R_\perp=1.75$ pMpc) and velocity offset of $\Delta V=692$ km s$^{-1}$.
QSOP2 is a quasar pair at $z=3.330$ and $z=3.309$, which is close to the edge of our redshift cut, with their angular separation $\Delta\theta=139$ arcseconds ($R_\perp=1.04$ pMpc) and velocity offset of $\Delta V=692$ km s$^{-1}$.
Note that only \citet{Fukugita04} has investigated the overdensity around $z>3$ quasar pairs before this study.
We search for their radio counterparts in the Faint Images of the Radio Sky at Twenty cm survey (FIRST 14Dec17 version; \cite{Becker95}) within 30 arcsecond radius, but find that none of them are detected.
There are $\sim30$ pairs in the whole HSC-Wide coverage (i.e., $1400$ deg$^2$).
The complete analysis of these pair environments will be done after the HSC-SSP survey is completed.
Note that our pair selection is incomplete for small-scale pairs due to the fiber collision limit of the BOSS survey ($\Delta \theta=62$ arcseconds).
We check whether high-redshift physical pairs previously identified in the literature (i.e., \cite{Schneider00, Hennawi10}) are covered in DR S16A, but find no suitable pairs in our redshift range.
There are several already-known pairs in the entire HSC-SSP survey regions, as we describe in Section~\ref{sec:discussion_highz}.

\subsection{Imaging data and method}\label{sec:z34_selection}
The $g$-dropout selection in this paper is the same as \citet{Toshikawa17}.
We first apply a magnitude cut of $i<25$ ($=i_\mathrm{lim,5\sigma}-1.4$) to measure the overdensity in a magnitude range satisfying high completeness.
Since this threshold corresponds to $\sim M^*+2$ at $z\sim4$ \citep{Bouwens07} where $M^*$ denotes the characteristic magnitude,
our density measurements are limited to the bright population.
In addition, we require significant detection in $r$- ($<r_\mathrm{lim,3\sigma}$) and $i$-bands ($<i_\mathrm{lim,5\sigma}$) to remove contaminants such as artificial and moving objects.
It is noted that the broad-band selection of Lyman break galaxies has a large uncertainty on their redshifts, corresponding to $\sim200$ pMpc in the line-of-sight direction, which is much larger than the projection direction.
Then, the following color cuts are applied.\\
If $g<g_\mathrm{lim,3\sigma}$,
\begin{eqnarray}
1.0&<&g-r \\
 -1.0&<&r-i<1.0 \\
1.5(r-i) &<& (g-r) - 0.8
\end{eqnarray}
, and if $g\geq g_\mathrm{lim,3\sigma}$,
\begin{eqnarray}
1.0&<&g_\mathrm{lim,3\sigma}-r\\
-1.0&<& r-i < 1.0\\
1.5(r-i) &<& (g_\mathrm{lim,3\sigma}-r) - 0.8.
\end{eqnarray}
Note that the observed magnitudes are corrected for the Galactic extinction.
Furthermore, we require that each source is de-blended with others ({\tt deblend\_nchild=0}) and
passes various photometric quality flags of the HSC-SSP
\footnote{Specifically, we use the following flags: {\tt centroid\_sdss\_\\
flags,flags\_pixel\_edge,flags\_pixel\_interpolated\_center,\\
flags\_pixel\_saturated\_center,flags\_pixel\_cr\_center,flags\\
\_pixel\_bad,flags\_pixel\_suspect\_center,cmodel\_flux\_flags}}.

Our overdensity measurements of $g$-dropout galaxies in the quasar pair fields are as follows.
First, we extract $g$-dropout galaxies within $2\times2$ deg$^2$ field centered on the pair in the projection plane.
Then, we set a square grid on the field at $0.6$ arcminute intervals to measure the number count of $g$-dropouts at each position
within a $1.8$-arcminutes aperture,
corresponding to the size of a typical proto-cluster at $z\sim4$ ($0.75$ pMpc).
We calculate the average and standard deviation of the $g$-dropout counts over the effective region (see Section~\ref{sec:eff_region}) inside the $2\times2$ deg$^2$ fields.
Blank grids where no galaxies exist inside the aperture are also masked out.
The total area effectively used in $2\times2$ deg$^2$ is $2.44$ deg$^2$ for QSOP1 and $2.19$ deg$^2$ for QSOP2, which are large enough to calculate the field number counts.
After deriving the significance map over the wide area, we zoom in on the pair vicinity of $12\times12$ arcmin$^2$ ($\sim5\times 5$ pMpc$^2$) to see its local overdensity significance.
The arbitrary zoom-in scale is chosen to be larger than the pair separation and thus enough to see the overdensity structure in the quasar pair fields.

\begin{table*}
  \tbl{Quasar pair sample at $3<z<4$}{%
  \begin{tabular}{ccccccccc}
  \hline              
  ID$^a$ &   R.A.$^b$ & Decl.$^c$ & redshift$^d$ & $i^e$& $\Delta\theta^f$& $R_\perp^g$ & $\Delta V^h$ \\ 
   &  (J2000) & (J2000) &  & [mag]&  [arcsec] & [pMpc] & [km\ s$^{-1}$]\\ 
    \hline
  QSOP1 & 22:14:52:49 & +01:11:19.9 & 3.585 &$21.167\pm0.002$& 240.6 &1.75 & 692\\
   &  22:14:58.38 & +01:07:36.1  & 3.574 &$20.380\pm0.001$& & &  \\
  QSOP2 &  16:14:47.39 & +42:35:25.2& 3.330 &$20.373\pm0.001$& 139.0 &1.04 & 1448\\
    &  16:14:51.35& +42:37:37.2  & 3.309 &$20.092\pm0.001$& & &  \\
  \hline
    \end{tabular}}\label{tab:QSOP_z34}
\begin{tabnote}
{\bf Notes:} $^a$ Pair ID. $^{b,c}$ HSC coordinates. $^d$ SDSS DR12Q visual redshift ({\tt Z\_VI}). $^e$  Extinction-corrected HSC-$i$ magnitude. $^f$ Angular separation in arcseconds. $^g$ Projected separation in physical scale. $^h$ Velocity offset of the pairs in km s$^{-1}$.  
\end{tabnote}
\end{table*}

\section{Result {\sc I}: $z>3$ Quasar Pair Environments}\label{sec:results_z34}
\subsection{Discovery of proto-clusters at $z=3.3$ \& $3.6$}
\begin{figure*}[tb]
\begin{minipage}{0.95\columnwidth}
 \begin{center}
  \FigureFile(83mm, 80mm){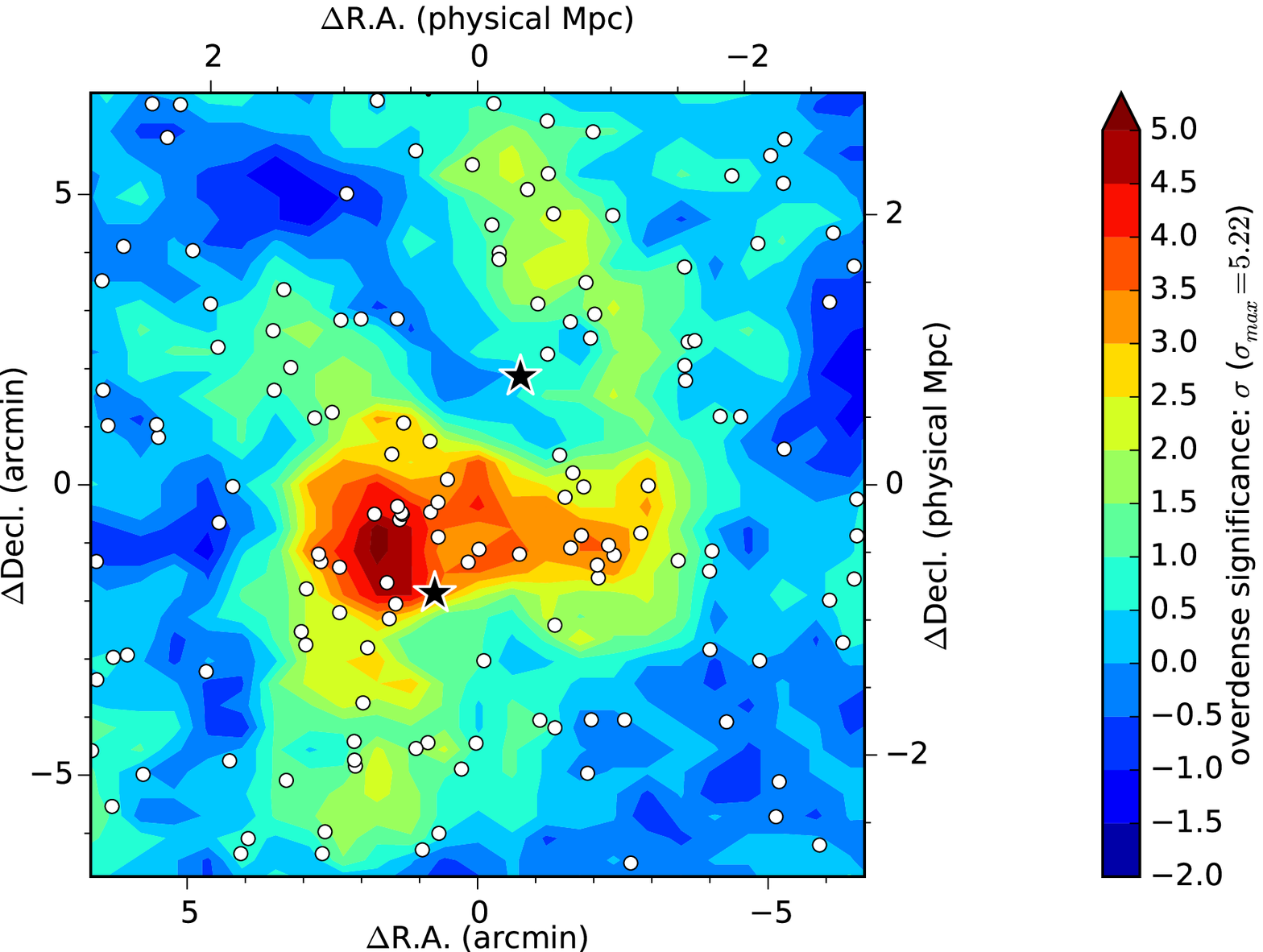} 
   \end{center}
    \end{minipage}
    \hspace{0.10\columnwidth}
\begin{minipage}{0.95\columnwidth}
 \begin{center}
  \FigureFile(83mm, 80mm){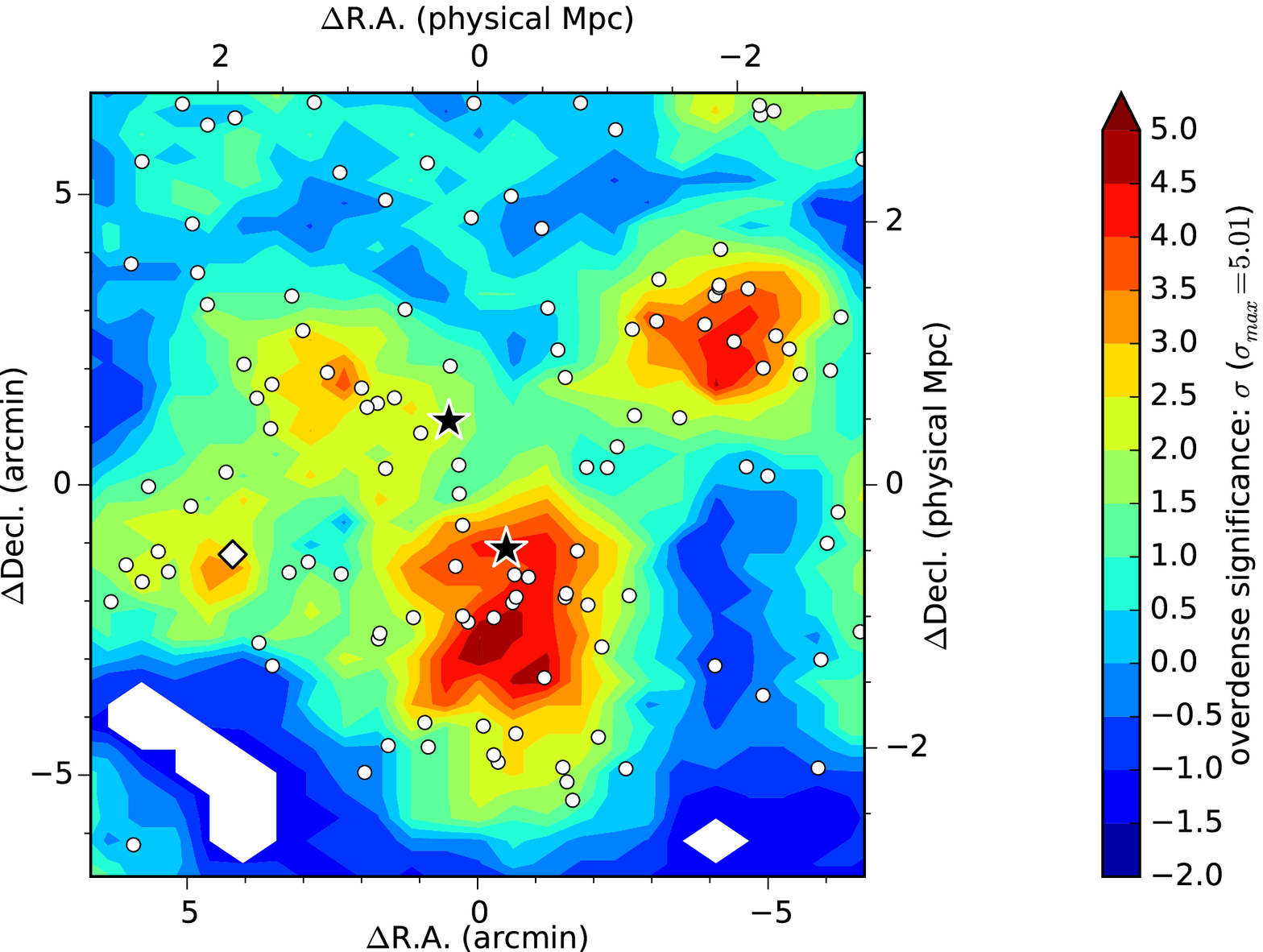} 
   \end{center}
    \end{minipage}
\caption{Overdensity profile of $g$-dropout galaxies ($12\times12$ arcminutes$^2$) around QSOP1 (left) and QSOP2 (right). 
The quasar pairs are shown in black stars.
The surrounding $g$-dropouts are shown in white circles. 
The $g$-dropout overdensity significance is shown in color contours. 
The region where no galaxies are found is masked.
A white diamond in the QSOP2 field shows the quasar candidate, HSCJ161506+423519 ($i=22.30$).
The physical scale from the pair center is also indicated.}\label{fig:fig1}
\end{figure*}

\begin{figure*}[htb]
 \begin{center} 
  \FigureFile(100mm, 45mm){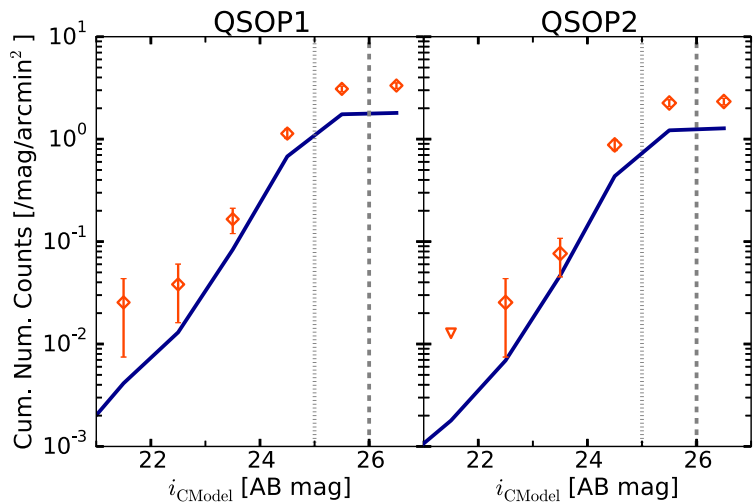} 
   \end{center}
\caption{
Cumulative $i$-band number counts of the $g$-dropouts within five arcminutes from the pair centers (left: QSOP1, right: QSOP2). 
In each panel, the open symbols show those of the pair field with Poisson error bars.
Magnitude bins with zero source count are shown in upside-down triangles.
The solid line shows the number counts of the field where the pair vicinity is excluded.
The $i=25$ and $i=26$ magnitude thresholds are shown in vertical dotted and dashed lines, respectively. 

}\label{fig:fig_ncount}
\end{figure*}

This section shows the result of the overdensity measurements for the two quasar pair fields at $z>3$.
We find significant overdensity in both pair fields with the peak significance $\sigma_\mathrm{peak}=5.22\sigma$ and $5.01\sigma$, which is summarized in Table~\ref{tab:z34result}.
Figure~\ref{fig:fig1} shows the overdensity profile of the pair fields where the color contours indicate the $g$-dropout overdensity significance based on the $g$-dropout counts.
The pair members are shown in stars and $g$-dropout galaxies are shown in circles.
QSOP1 field shows a filament-like structure in the westward direction and QSOP2 field shows a core-like structure with several smaller density peaks in the vicinity.
The four quasars themselves are not selected in our dropout selection due to their relatively blue $g-r$ colors ($0.3\leq g-r\leq0.9$), 
which could be explained by the intrinsic power-law shape of the quasar continuum declining toward longer wavelength.
The cumulative number counts of the $g$-dropouts within five arcminutes from the pair centers are shown in Figure~\ref{fig:fig_ncount} with open symbols, compared with those of the $2\times2$ deg$^2$ general fields used for the density measurements
\footnote{The pair vicinity is excluded in the calculation of the field number counts.}.
Overall, the number counts are roughly twice as high in all (non-zero) magnitude bins even within such a large projected area corresponding to $\sim2$ pMpc radius.
Specifically, there are two bright dropouts at $21.2<i<21.4$ in the QSOP1 field, which is six times higher density than the general fields.
Although the redshift uncertainty in our dropout selection is large, our result strongly suggests that these pairs are associated with massive environments, namely proto-clusters.

Indeed, QSOP1 and QSOP2 fields are part of the HSC proto-clusters cataloged in \citet{Toshikawa17}.
The two fields also have high sigma peaks over 4$\sigma$ in their measurements ($4.8 \sigma$ and $4.0 \sigma$, respectively), which are likely to evolve in massive clusters in the local universe with the descendant halo mass $M_\mathrm{halo,z=0}>10^{14} M_\odot$.
It is notable that these two fields are not the richest among their HSC proto-clusters, which could suggest that even quasar pairs do not trace the most massive haloes.
The reason the significance in our measurements is slightly higher than their measurements is the difference of the area used for the significance measurements.
In \citet{Toshikawa17}, they measure the average and standard deviation of $g$-dropout counts over the whole S16A HSC-Wide area, deriving the average  $N_\mathrm{ave, Wide}=6.39\pm 3.24$. 
Meanwhile, we measure the same quantities in an area approximately corresponding to the HSC field-of-view around the pairs.
Therefore, their overdensity measurements reflect not only intrinsic galaxy density distribution but also different completeness over the large area due to different observation depth taken in various seeing conditions and different times of being covered while dithering.
Indeed, the peak significance of the two pair fields is smaller in our measurements, and both fields, especially QSOP2 at the edge of the W-HECTMAP, have smaller $N_\mathrm{ave}$ than their measurements.

For the two proto-clusters, we find that local peaks are close to one of the pair members, but not to the other one (see Table~\ref{tab:QSOP_z34}).
For QSOP1, the significance just above the two quasars are not high and, in particular, HSCJ221452+011119 (the upper one in Figure~\ref{fig:fig1}) is at the outskirt of the overdensity profile.
This is also the case for QSOP2, as HSCJ161451+423737 (the upper one) is at the gap of density peaks.
Nevertheless, there are only six out of $151$ quasars in \citet{Uchiyama17} measurements with which $>4\sigma$ overdensity regions are associated within three arcminutes from the HSC proto-clusters, and intriguingly, three of them are the quasar pairs: the two QSOP1 quasars and one QSOP2 quasar (HSCJ161447+423525).
Therefore, our result may indicate that pairs of quasars are likely related to rich environments,
but they do not emerge at the central peak of galaxy density.
\begin{table}
  \tbl{Overdensity significance around two $z>3$ quasar pairs based on $i<25$ $g$-dropouts}{%
  \begin{tabular}{ccccc}
      \hline
  ID &   $\sigma_\mathrm{peak}^a$ & $\sigma_\mathrm{Q1}^b$ & $\sigma_\mathrm{Q2}^c$  &$(N_\mathrm{ave}\pm\sigma_\mathrm{STD})^d$   \\ 
      \hline
  QSOP1 & 5.22& -0.02& 3.30& 5.80 $\pm$2.91 \\
  QSOP2 & 5.01& 4.21& 2.23& 4.40 $\pm$2.52\\
      \hline
    \end{tabular}}\label{tab:z34result}
\begin{tabnote}
{\bf Notes:} $^a$ Overdensity significance. $^{b,c}$ Significance above each quasar. The former and latter quasars in Table~\ref{tab:QSOP_z34} are denoted as Q1 and Q2, respectively.
$^d$ Average number and standard deviation ($=\sigma_\mathrm{STD}$) of $g$-dropouts within a $1.8'$-radius aperture.
\end{tabnote}
\end{table}

\begin{figure*}[bt]
\begin{minipage}{0.95\columnwidth}
 \begin{center}
  \FigureFile(83mm, 80mm){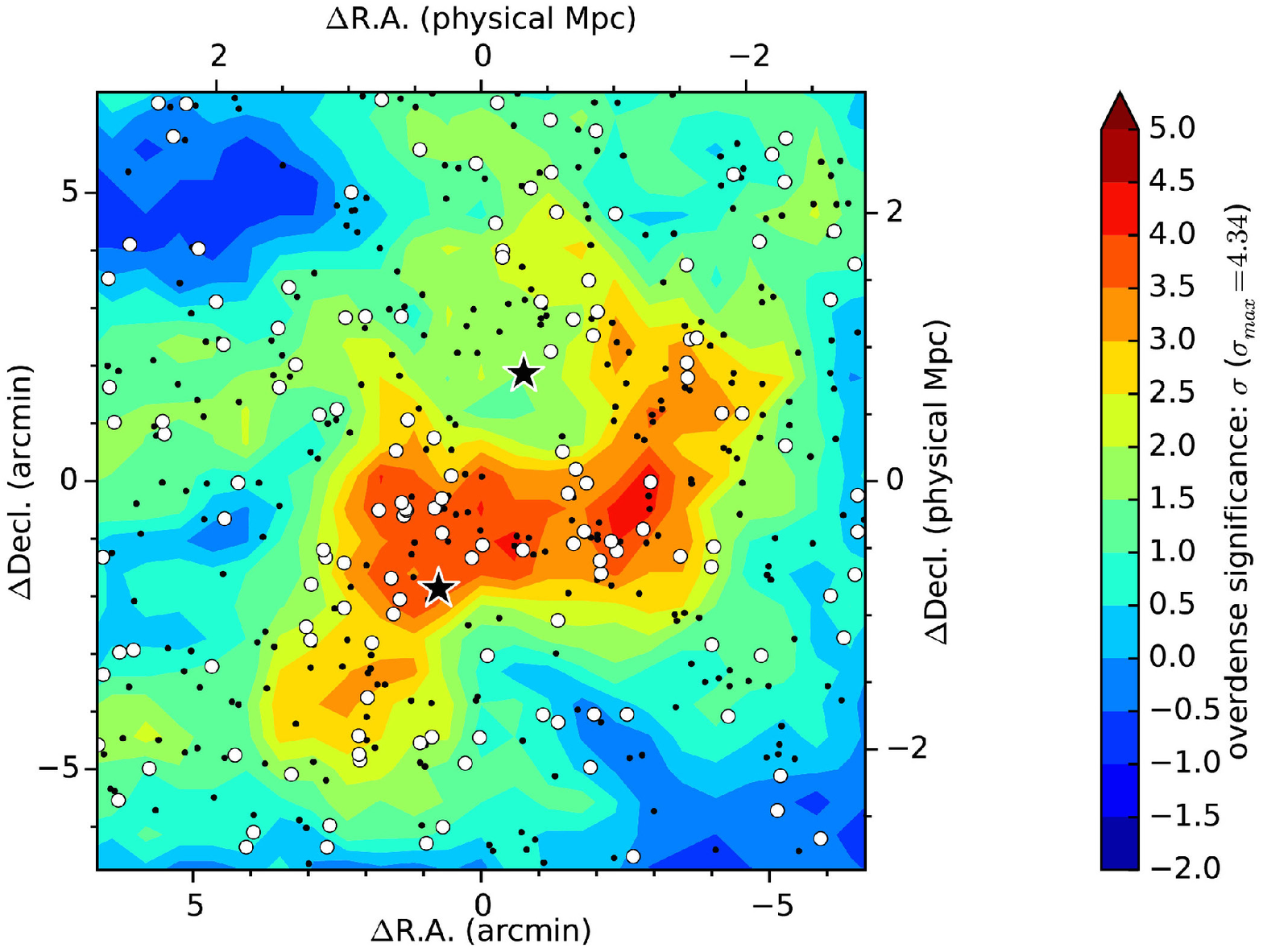} 
   \end{center}
    \end{minipage}
    \hspace{0.10\columnwidth}
\begin{minipage}{0.95\columnwidth}
 \begin{center}
  \FigureFile(83mm, 80mm){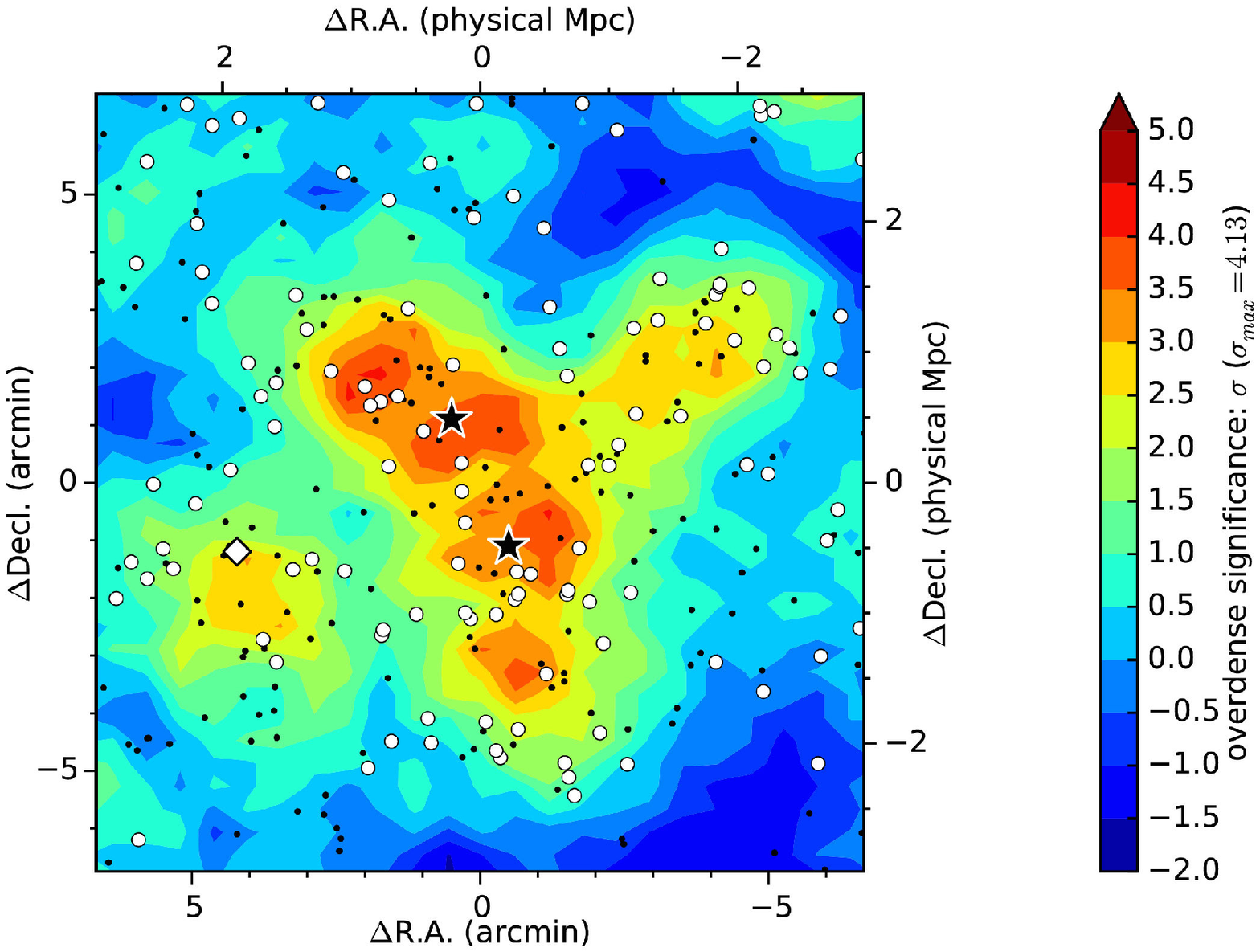} 
   \end{center}
    \end{minipage}

\caption{The same significance maps for QSOP1 and QSOP2, but with fainter $g$-dropouts down to $i=26.0$.
The black dots show the faint ($25\leq i<26$) dropouts. 
The overdensity significance are measured with all $i< 26.0$ dropouts.}\label{fig:fig2}
\end{figure*}

To further investigate the overdensity structure, the overdensity profiles of the same two fields are measured again including fainter $g$-dropouts down to the approximate $5\sigma$ limiting magnitude.
Figure~\ref{fig:fig2} shows the local significance maps of $g$-dropouts down to $i<26$ ($\sim i_\mathrm{lim,5\sigma}$), for which only the $i$-band magnitude cut is loosened by one magnitude in the original dropout selection criteria.
The circles and stars are the same as Figure~\ref{fig:fig1} and the black dots show the $g$-dropouts with $25\leq i<26$.
In the two significance maps, the overdensity structures become more extended and centered around the pairs in Figure~\ref{fig:fig2}.
While the completeness falls down from $i>25$ as is also indicated in Figure~\ref{fig:fig_ncount}, the profiles would trace real structures assuming that the completeness is independent of local positions.
Therefore, we conclude that these proto-clusters have filament-like extended structure and likely host the luminous quasars pairs inside.
We discuss the interpretation of our results in Section~\ref{sec:discussion_highz} comparing with single quasar environments.

\subsection{A faint quasar candidate in the pair fields}\label{sc:bright_gal}
We inspect the possibility that the two proto-clusters we find host other quasars fainter than the BOSS depth limit.
Here, we assume a pure point source morphology and search for faint quasar candidates among the $g$-dropouts in the two pair fields. 
For this purpose, a shape parameter included in the HSC-SSP dataset is used, which gives a rough shape measurement based on the second moment of the object image.
We take the $i$-band shape parameter, since the $i$-band observation in the HSC-SSP is executed in good seeing conditions (typically $0.56$ arcsecond).
Using the ratio of the $g$-dropout moment ({\tt ishape\_sdss}, $I_{ij}$) over the PSF moment ({\tt ishape\_sdss\_psf}, $\psi_{ij}$),
\citet{Akiyama17} show that, through their $z\sim4$ quasar selection, 
$i<23$ point sources are extracted with $>80\%$ completeness using their criteria: $I_{xx}/\psi_{xx}<1.1\ \wedge\ I_{yy}/\psi_{yy}<1.1$.
We apply the same criteria for our $g$-dropouts down to $i<23$ in the QSOP1 and QSOP2 fields.
Among five $g$-dropouts with $i<23$ in the QSOP1 and two in QSOP2, we find that the brightest dropout in the QSOP2 field, HSCJ161506+423519 ($i=22.30$) is a point source with its shape parameters $I_{xx}/\psi_{xx}=1.00$ and $I_{yy}/\psi_{yy}=1.01$, as is also shown in Table~\ref{tab:Faintqso}.
This quasar candidate is shown in a diamond in Figure~\ref{fig:fig1} and \ref{fig:fig2}.
This source is $\sim2$ pMpc away from the two QSOP2 quasars, and at the center of a small density peak.
Therefore, it is likely that three QSOP2 quasars cluster in $2$ pMpc scale, embedded in a large proto-cluster at $z\sim3.3$,
while the faint quasar candidate could be a foreground or background quasar independent of the pair fields.
We search for the radio counterpart of this quasar candidate with the FIRST survey, but do not find any source within $30$ arcsecond from the optical position.

\begin{table*}
  \tbl{The quasar candidate at $z\sim3.3$ associated with the QSOP2: HSCJ161506+423519}{%
  \scalebox{0.85}{
  \begin{tabular}{cccccccccccc}
  \hline    \hline              
    R.A. & Decl. & $g$ & $r$&$i$& $z$& $y$ & $I_{xx}^*$ & $I_{yy}^*$&$(I_{xx}/\psi_{xx})^\dagger $ & $(I_{yy}/\psi_{yy})^\dagger$ \\ 
     (J2000) & (J2000) &  [mag]& [mag]& [mag]& [mag] & [mag] &[arcsec$^2$] & [arcsec$^2$] & & \\ 
    \hline 
  16:15:06.24 & +42:35:19.4 & $24.414\pm0.033$ &$22.732\pm0.008$&$22.272\pm0.005$&$22.123\pm0.012$ &$22.076\pm0.025$ &$0.0442$ & $0.0426$&$1.00$& $1.01$&\\
  \hline
\end{tabular}}}\label{tab:Faintqso}
\begin{tabnote}
{\bf Notes:} The HSC magnitudes are extinction-corrected.\\
 $^*$ The practical adaptive momentum of the object. \\
 $^\dagger$ ratio of the object momentum over that of the PSF model: $I/\psi$.
Note that point sources can be extracted with $>80\%$ completeness down to $i=23$ with $(I_{xx}/\psi_{xx})<1.1 \wedge (I_{yy}/\psi_{yy})<1.1$ (see Section~2.2 of \cite{Akiyama17}).

\end{tabnote}
\end{table*}

\section{Quasar Pairs at $z\sim1$}\label{sec:lowz_pair}
In the previous sections, we report that two quasar pairs at high redshift do trace proto-clusters.
On the other hand, previous studies show that quasar pairs at low redshift ($z<1$)  
do not always reside in dense environments.
A recent study by \citet{Song16} shows that $z\sim1$ single quasar environments have a slight tendency toward high density regions,
while the enhancement of the quasar density is weaker than expected from a proportional relation of the galaxy density.
From this section, we extend the redshift range down to $z\sim1$ to compare with the high-redshift pairs and also with single quasars at the same redshift.
\subsection{$z\sim1$ quasar pair selection}
\subsubsection{BOSS pairs}
\begin{figure}[h]
 \begin{center}
  \FigureFile(80mm, 70mm){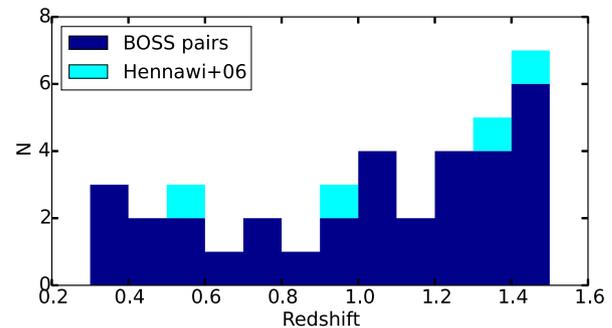} 
   \end{center}
\caption{Redshift distribution of the $z<1.5$ pairs.
$33$ BOSS quasar pairs ($R_\perp<2$ pMpc and $\Delta V<2300$ km s$^{-1}$) is shown in dark blue.
The four pairs from \citet{Hennawi06} are shown in cyan (Sec.~\ref{sec:binary}). 
The pair coordinates, redshift, separation are summarized in Table~\ref{tab:tab_lowz_b}.
}\label{fig:lowz_hist}
\end{figure}

We extract our sample of low-redshift quasar pairs from the SDSS DR12Q catalog as follows.
First, the selection area is limited to the $\sim 121$ deg$^2$ of the S16A effective area.
Second, we limit the redshift range to $z<1.5$, over which the photometric redshift estimate with the five-band photometry of the HSC has a large scatter and a high contamination rate \citep{Photozpaper}.
After applying the BOSS redshift flag ({\tt ZWARNING=0}), we apply our definition of $z\sim1$ quasar pairs:
two quasars within projected separation $R_\perp<2$ ($=1.4\ h^{-1}$) pMpc and velocity offset $\Delta V<2300$ km s$^{-1}$.
The maximum projected separation corresponds to the size of a $z\sim1$ proto-cluster, the descendant halo mass of which is $M_{\mathrm{halo,}z=0}\sim10^{14}M_\odot$ \citep{Chiang13}.
It is chosen to select pairs with comparable separation to the two $z>3$ pairs in this paper.
The maximum velocity difference takes into account redshift uncertainty, peculiar velocity, and physical separation of $<2$ pMpc. 
At this stage, we select $38$ pairs.
We further require that the following positions and areas are within the S16A effective area to exclude insufficient fields for the overdensity measurements: 
i) at the position of the quasars, ii) at the pair center, iii) over 70\% of the $2\times2$ deg$^2$ area centered on the pair, and iv) over 80\% of the pair vicinity ($8\times 8$ arcmin$^2$).
Finally, $33$ pairs at $0.33<z<1.49$ ($z\sim1.02$ on average) are extracted in the S16A area.
Their redshift distribution is shown in Figure~\ref{fig:lowz_hist}.
We find that J020332.82-050944.5 at $z=1.353$ is double-counted, having two companions nearby.
Since the projection separation of the two companions are over the cluster scale (i.e., $>2$ pMpc), we treat the two pairs individually in the following analysis.
It is noted that whether these three quasars are considered as two pairs or a triplet does not affect our final result in Section~\ref{sec:lowz_pair_res}.
In the search of $0.5<z<3$ small-scale quasar pairs with the SDSS \citep{Hennawi06}, two quasars with $R_\perp<1$ $h^{-1}$ pMpc and $\Delta V<2000$ km s$^{-1}$ separation are assumed to be physically associated binaries.
On the other hand, this study loosens the pair selection criteria as we recognize a cluster-scale association of two quasars as a pair.
While we have four small-scale ($R_\perp<1$ pMpc) pairs from the BOSS catalog, it should be noted that our selection is not complete, since a complete search of such sub pMpc-scale pairs requires a dedicated spectroscopic campaign.
The detailed information of the pairs such as their coordinates, redshift, and pair separation are given in Table~\ref{tab:tab_lowz}.
Note that all $z\sim1$ pairs are in the W-XMMLSS region.
We find that most of them are flagged as the BOSS ancillary program targets ({\tt ANCIALLY\_TARGET2}).

\begin{longtable}{lcccccc}
\caption{Quasar pairs at $z\sim1$ extracted from the BOSS DR12Q with their overdensity significance within $2$ arcminutes}\label{tab:tab_lowz}
  \hline\hline              
  ID$^a$ &  redshift$^b$ & $i^c$& $\Delta\theta$& $R_\perp$ & $\Delta V$ & $\sigma_\mathrm{peak, 2'}^d$\\ 
    &  & [mag]&  [arcsec] & [pMpc] & [km\ s$^{-1}$] & \\ 
    \hline 
    \endfirsthead
    \hline\hline
  ID$^a$ &  redshift$^b$ & $i^c$& $\Delta\theta$& $R_\perp$ & $\Delta V$ & $\sigma_\mathrm{peak, 2'}^d$\\ 
    &  & [mag]&  [arcsec] & [pMpc] & [km\ s$^{-1}$] & \\ 
    \hline
    \endhead
    \hline
   \endfoot
       \hline
    \multicolumn{7}{l}{{\bf Notes:} $^a$ DR12Q ID. $^b$ DR12Q redshift ({\tt Z\_VI}). $^c$ SDSS-$i$ PSF magnitude. $^d$ Peak significance}\\
     \multicolumn{7}{l}{within two arcminutes around the pairs.}\\
      \multicolumn{7}{l}{$^*$ The quasar having two companions.  $^\dagger$ Small-scale pairs with $R_\perp<1$ pMpc.}
    \endlastfoot    
SDSSJ020257.39-051225.4&0.512&21.13 $\pm$ 0.09&251.3&1.55&504&2.08\\
SDSSJ020313.15-051057.6 & 0.514 & 21.30 $\pm$ 0.10 &  &  &  & \\
SDSSJ020320.47-050933.8 & 1.3526 & 21.18  $\pm$ 0.09& 184.7 & 1.55 & 40 & 1.79\\
SDSSJ020332.82-050944.5$^*$ & 1.3529 & 21.10  $\pm$ 0.09&  &  &  & \\
SDSSJ020332.82-050944.5$^*$ & 1.3529 & 21.10  $\pm$ 0.09& 182.7 & 1.54 & 1057 & 5.18\\
SDSSJ020341.74-050739.5 & 1.345 & 19.69  $\pm$ 0.05&  &  &  & \\
SDSSJ020334.58-051721.3$^\dagger$ & 1.399 & 21.30  $\pm$ 0.11& 99.9  &  0.84 & 130 & 1.62\\
SDSSJ020336.45-051545.4$^\dagger$& 1.400 & 20.42  $\pm$ 0.02&  &  &  & \\
SDSSJ020411.47-051032.7 & 0.326 & 19.86  $\pm$ 0.04& 314.3 & 1.49 & 1261 & 1.39\\
SDSSJ020423.94-050619.6 & 0.332 & 19.74  $\pm$ 0.03&  &  &  & \\
SDSSJ020442.23-051041.3 & 1.329 & 20.90  $\pm$ 0.07& 117.5 & 0.99 & 374 & 0.92\\
SDSSJ020448.12-050923.4 & 1.326 & 22.77  $\pm$ 0.32&  &  &  & \\
SDSSJ020645.57-044511.8 & 1.410 & 19.53  $\pm$ 0.07& 226.6 & 1.91 & 2262 & 4.57\\
SDSSJ020700.09-044406.8 & 1.428 & 20.72  $\pm$ 0.08&  &  &  & \\
SDSSJ020659.51-042343.3 & 0.732 & 21.21  $\pm$ 0.08& 173.0 & 1.26 & 587 & 2.94\\
SDSSJ020709.83-042501.5 & 0.729 & 21.32  $\pm$ 0.08&  &  &  & \\
SDSSJ021105.40-051424.0$^\dagger$ & 1.078 & 21.37  $\pm$ 0.08& 37.0 & 0.30 & 680 & 4.56\\
SDSSJ021107.88-051425.6$^\dagger$ & 1.083 & 20.04  $\pm$ 0.04&  &  &  & \\
SDSSJ021335.21-055002.7 & 1.2219 & 18.86 $\pm$ 0.02 & 173.4 & 1.44 & 18 & 2.41\\
SDSSJ021339.82-055241.9 & 1.2220 & 20.56  $\pm$ 0.05&  &  &  & \\
SDSSJ021425.44-035631.7 & 1.423 & 20.63  $\pm$ 0.05& 137.5 & 1.16 & 356 & 1.12\\
SDSSJ021434.30-035555.3 & 1.426 & 20.88  $\pm$ 0.05&  &  &  & \\
SDSSJ021436.80-045150.4 & 1.107 & 20.60  $\pm$ 0.21& 156.6 & 1.28 & 1091 & 1.34\\
SDSSJ021444.98-045012.6 & 1.115 & 21.50  $\pm$ 0.10&  &  &  & \\
SDSSJ021448.84-040601.7 & 0.4447 & 20.22  $\pm$ 0.04& 293.4 & 1.68 & 74 & 1.83\\
SDSSJ021508.19-040513.6 & 0.4451 & 20.07  $\pm$ 0.03&  &  &  & \\
SDSSJ021606.59-040508.4 & 1.1447 & 20.19  $\pm$ 0.13& 147.3 & 1.21 & 26 & -0.11\\
SDSSJ021614.95-040626.3 & 1.1449 & 21.40  $\pm$ 0.08&  &  &  & \\
SDSSJ021610.64-045229.8 & 0.9570 & 21.35  $\pm$ 0.07& 240.7 & 1.91 & 28 & 3.43\\
SDSSJ021626.53-045308.3 & 0.9568 & 21.17  $\pm$ 0.07&  &  &  & \\
SDSSJ021650.21-040142.6 & 1.031 & 20.96  $\pm$ 0.06& 218.3 & 1.76 & 154 & 2.85\\
SDSSJ021659.43-035853.4 & 1.030 & 21.77  $\pm$ 0.11&  &  &  & \\
SDSSJ021710.20-034101.2 & 1.425 & 20.85  $\pm$ 0.07& 178.1 & 1.50 & 296 & 1.37\\
SDSSJ021718.77-033857.6 & 1.427 & 21.59  $\pm$ 0.13&  &  &  & \\
SDSSJ021756.83-035316.6 & 0.7511 & 20.53  $\pm$ 0.04& 231.2 & 1.70 & 92 & 4.20\\
SDSSJ021806.76-035019.5 & 0.7505 & 21.72  $\pm$ 0.13&  &  &  & \\
SDSSJ021757.23-050216.3 & 1.088 & 20.47  $\pm$ 0.04& 237.0 & 1.93 & 1081 & 6.46\\
SDSSJ021809.48-045945.8 & 1.095 & 21.60  $\pm$ 0.10&  &  &  & \\
SDSSJ021809.55-050200.3 & 1.277 & 21.22  $\pm$ 0.07& 128.3 & 1.07& 1290 & 3.96\\
SDSSJ021817.20-050258.7 & 1.287 & 20.26  $\pm$ 0.04&  &  &  & \\
SDSSJ022024.49-040017.2 & 0.812 & 21.58  $\pm$ 0.11& 228.2 & 1.72 & 1723 & 0.75\\
SDSSJ022032.41-040332.2 & 0.822 & 20.77  $\pm$ 0.06&  &  &  & \\
SDSSJ022125.05-055638.0$^\dagger$ & 0.585 & 20.59  $\pm$ 0.04& 150.4 & 0.99 & 893 & 2.98\\
SDSSJ022128.77-055857.7$^\dagger$ & 0.580 & 19.81  $\pm$ 0.02&  &  &  & \\
SDSSJ022226.84-041313.4 & 1.486 & 21.73  $\pm$ 0.11& 189.7 & 1.60 & 1193 & 1.73\\
SDSSJ022237.88-041140.3 & 1.496 & 21.52  $\pm$ 0.09&  &  &  & \\
SDSSJ022248.98-044824.6 & 1.419 & 20.81  $\pm$ 0.06& 200.7 & 1.69 & 263 & 1.97\\
SDSSJ022253.17-044513.9 & 1.421 & 21.17  $\pm$ 0.09&  &  &  & \\
SDSSJ022534.82-042401.6 & 0.920 & 21.06  $\pm$ 0.08& 152.8 & 1.20 & 222 & 5.90\\
SDSSJ022537.16-042132.9 & 0.921 & 19.32  $\pm$ 0.03&  &  &  & \\
SDSSJ022542.41-051452.4 & 1.258 & 21.35  $\pm$ 0.11& 194.8 & 1.63 & 281 & 1.82\\
SDSSJ022554.86-051354.6 & 1.256 & 20.08  $\pm$ 0.04&  &  &  & \\
SDSSJ022550.97-040247.4 & 1.448 & 21.14  $\pm$ 0.08& 150.1 & 1.27 & 799 & 2.25\\
SDSSJ022552.15-040516.5 & 1.441 & 21.30  $\pm$ 0.08&  &  &  & \\
SDSSJ022855.35-051130.6 & 0.366 & 18.97  $\pm$ 0.05& 200.5 & 1.02 & 96 & 0.78\\
SDSSJ022855.95-051450.8 & 0.365 & 19.96  $\pm$ 0.06&  &  &  & \\
SDSSJ022916.82-044600.7 & 0.612 & 20.35  $\pm$ 0.04& 193.9 & 1.31 & 455 & 1.60\\
SDSSJ022928.73-044444.0 & 0.610 & 20.53  $\pm$ 0.05&  &  &  & \\
SDSSJ023035.82-052603.2$^\dagger$ & 0.364 & 19.68  $\pm$ 0.03& 153.2 & 0.78 & 318 & 1.84\\
SDSSJ023038.66-052336.0$^\dagger$ & 0.363 & 19.96  $\pm$ 0.03&  &  &  & \\
SDSSJ023231.43-053655.9 & 1.098 & 20.92  $\pm$ 0.07& 165.5 & 1.35 & 408 & 3.18\\
SDSSJ023238.46-053903.9 & 1.101 & 21.01  $\pm$ 0.07&  &  &  & \\
SDSSJ023323.39-042803.0 & 1.238 & 21.41  $\pm$ 0.08& 118.1 & 0.98 & 362 & 2.42\\
SDSSJ023331.24-042815.2 & 1.241 & 20.76  $\pm$ 0.06&  &  &  & \\
SDSSJ023328.44-054604.4$^\dagger$ & 0.494 & 20.31  $\pm$ 0.03& 41.1 & 0.25 & 170 & 2.46\\
SDSSJ023331.05-054550.9$^\dagger$ & 0.493 & 18.45  $\pm$ 0.02&  &  &  &\\
\end{longtable}

\subsubsection{Pairs in the literature}\label{sec:binary}
To complement the sub pMpc-scale quasar pairs lacking in the BOSS pair selection, spectroscopically confirmed small-scale pairs identified in the literature are added to our sample.
Such pairs are usually identified as byproducts of lensed quasar searches.
We look for spectroscopically confirmed binary quasars within the DR S16A coverage referring to the following literature: \citet{Djorgovski91, Myers08, Hennawi06, Kayo12, Inada12, More16, Eftekharzadeh17}.
Following the same selection procedure applied to the BOSS quasars, we are able to measure the galaxy overdensity around four small-scale quasars originally identified in \citet{Hennawi06}.
Therefore, the total number of quasar pairs including pairs from literature is $37$, nine of which are $R_\perp<1$ pMpc pairs.
The redshifts of the additional pairs are shown in cyan in Figure~\ref{fig:lowz_hist}, and their properties are listed in Table~\ref{tab:tab_lowz_b}.
The projected separation of the four pairs is almost comparable to the BOSS pairs, but
SDSSJ1152-0030 ($z\sim0.55$) has the smallest projected separation $R_\perp=0.13$ pMpc ($\Delta\theta=29.3$ arcseconds) among our sample.

\begin{table*}[htb]
  \tbl{ Quasar pairs at $z<1.5$ from \citet{Hennawi06}}{%
  \begin{tabular}{lcccccccc}
  \hline              
  ID &   redshift & $i$& $\Delta\theta$& $R_\perp$ & $\Delta V$ & $\sigma_\mathrm{peak, 2'}$\\ 
   &  & [mag]&  [arcsec] & [pMpc] & [km\ s$^{-1}$] & \\ 
    \hline 
  SDSSJ1152-0030A &0.550 &18.80 $\pm$ 0.02&29.3&0.19 &740 &5.07\\
  2QZJ1152-0030B &0.554 &19.99 $\pm$ 0.03&&&&\\
  2QZJ1209+0029A &1.319 & 20.26 $\pm$ 0.04& 165.8 &1.39&340&0.94 \\
  2QZJ1209+0029B &1.322 &20.76 $\pm$ 0.05&&&&\\
  2QZJ1411-0129A &0.990 & 19.85 $\pm$ 0.04& 152.5 &1.22&1820&1.14 \\
  2QZJ1411-0129B &0.978 &19.75 $\pm$ 0.03&&&&\\
  2QZJ1444+0025A &1.460 & 20.34 $\pm$ 0.04 & 113.3 &0.96&1950&0.43 \\
  2QZJ1444+0025B &1.444 &20.35 $\pm$ 0.04&&&&\\
  \hline
\end{tabular}}\label{tab:tab_lowz_b}
\begin{tabnote}
{\bf Notes:} ID, redshift, separation are derived from their measurements converted with the cosmology we adopt.
The $i$-band magnitudes are the extinction-corrected PSF magnitudes derived from the SDSS DR12.
In the last column, we report the peak significance within two arcminutes around the pair center.
\end{tabnote}
\end{table*}

\subsection{$z\sim1$ single quasars}
To compare with the $z\sim1$ pairs, we also measure the overdensity around single quasars at $0.9<z<1.1$ in the W-XMMLSS region.
To extract only isolated quasars, we require that there is no neighborhood quasar at $R_\perp<4$ pMpc and $\Delta V<3000$ km s$^{-1}$, in addition to the BOSS redshift flag.
We set this boundary larger than the maximum pair separation to remove companion quasars associated with quasar pairs.
We also check to see whether their fields are suitable for overdensity measurements using the same criteria applied to the quasar pair fields.
As a result, $127$ isolated quasars at $z\sim1$ are extracted, the sample size of which is large enough to statistically compare with the pair environments.

\subsection{$z\sim1$ galaxy selection}\label{sec:lowz_selection}
We measure the galaxy overdensity of the $37$ pair and $127$ single quasar fields at $z\sim1$, using a photometric redshift catalog available in the DR S16A ({\tt Mizuki}, \cite{Tanaka15}) derived from the HSC-SSP survey.
The accuracy of the photometric redshift ($z_\mathrm{phot}$) with respect to the spectroscopic redshift ($z_\mathrm{spec}$) is often characterized by two conventional quantities.
The scatter is denoted as
\begin{equation}
\sigma_\mathrm{conv}\equiv 1.48\times MAD\left(\frac{z_\mathrm{phot}-z_\mathrm{spec}}{1+z_\mathrm{spec}}\right)
\end{equation}
, where MAD stands for median absolute deviation.
The outlier rate is denoted as 
\begin{equation}
f_\mathrm{outlier, conv}\equiv\frac{N(\frac{\mid z_\mathrm{phot}-z_\mathrm{spec} \mid}{1+z_\mathrm{spec}}> 0.15)}{N_\mathrm{total}}
\end{equation}
, where the denominator stands for the total number of test samples and the numerator stands for the number of outliers.
In the S16A dataset, the scatter and the outlier rate at $z\sim1$ is $\sigma_\mathrm{conv}\sim0.05$ and $f_\mathrm{outlier}\sim0.1$ under a moderate seeing condition ($0.7$ arcsecond).

We extract surrounding galaxies within $2\times 2$ deg$^2$ centered on the pairs and single quasars using the following criteria:
\begin{eqnarray}
i&<&24\\
r&<&r_\mathrm{lim,5\sigma}\\
\frac{\mid z_\mathrm{med} - z_\mathrm{QSOP} \mid}{1+z_\mathrm{QSOP}}&<&0.05 \label{eq:pz}
\end{eqnarray}
, where we use the median redshift $z_\mathrm{med}$ defined as $\int_0^{z_{med}} P(z) dz = 0.5$ as a photometric redshift. 
$z_\mathrm{QSOP}$ is the average redshift of the two quasars in pair.
The magnitude cut in $i$-band ($\sim i_\mathrm{lim, 5\sigma}-2$) is to select galaxies with reliable classification and without significant contamination such as higher-redshift ($z>3$) galaxies.
It is noted that we select galaxies brighter than $\sim M^*+2$ at $z\sim1$ based on the $M^*$ calculation in \citet{Boris07}.
The scatter and outlier plots as functions of $i$-band magnitudes are given in \citet{Photozpaper}.
We pick up galaxies at redshift within five percent of the pair redshift using the third criterion (Eq.~\ref{eq:pz}).
This limit is selected to match the scatter of the photometric redshift (i.e., $\sigma_\mathrm{conv}$).
Furthermore, quality flag cuts for the photometric redshift are applied\footnote{
Specifically, we use {\tt photoz\_prob\_star}$<0.1$ for removing Galactic stars, {\tt reduced\_chisq}$<10$ for removing sources with unusual optical SEDs, and {\tt photoz\_conf\_median}$>0.2$ for removing sources with flat $P(z)$ distribution. 
} in addition to the same photometry flags in Section~\ref{sec:z34_selection}.
For the selected galaxies, we measure the local ovedensity significance in the pair fields following the same procedure for the high redshift pairs.
We first calculate the galaxy density map at each grid in $2\times2$ deg$^2$ around the pairs using the $1.8$ arcminutes aperture, and then measure the local peak of the overdensity significance within two arcminutes from the pair center.
At $z\sim1$, its diameter corresponds to $\sim2$ pMpc, comparable to the maximum pair separation.
We use the peak significance to quantify the pair environments and compare it with that of other environments such as galaxies and isolated quasars .

\subsection{Random fields around $z\sim1$ galaxies}\label{sec:lowz_random}
To compare with the pair and single quasar fields, we compute the peak significance distribution around randomly selected $z\sim1$ galaxies.
For this purpose, the random fields should be independent of the quasar presence.
We make use of the $2\times 2$ deg$^2$ fields around the $127$ single quasars at $0.9<z<1.1$, and galaxies selected by the photometric redshift selection in Section~\ref{sec:lowz_selection} (i.e., $i<24$).
We assume no significant redshift evolution of the surface galaxy density within $0.3<z<1.4$.
In each of the $127$ fields, we randomly pick ten galaxies removing the masked regions, the central $30\times30$ arcminutes$^2$ around quasars, and $<15$ arcminutes at the edges.
We then measure the significance peak within the two arcminutes radius to quantify the overdensity of the selected regions.
We require that there is no overlap between each aperture centered at the randomly selected galaxies to ensure the independence of the selected regions.
A certain amount of the $2\times2$ deg$^2$ fields overlaps, but this does not affect the randomness of the galaxy selection unless the selected galaxies are in vicinity of others.
After examining the randomly selected positions of the galaxies to check for any overlaps of the two-arcminutes aperture with others, 
we finally derive the overdensity significance distribution around $z\sim1$ galaxies from $849$ random fields, which is sufficient to compare the significance distribution with the quasar fields.

\section{Result {\sc II}: $z\sim1$ Quasar Pair Environments}\label{sec:lowz_pair_res}
\subsection{Significance distribution}
\begin{figure*}[ht]
 \begin{center}
  \FigureFile(160mm, 80mm){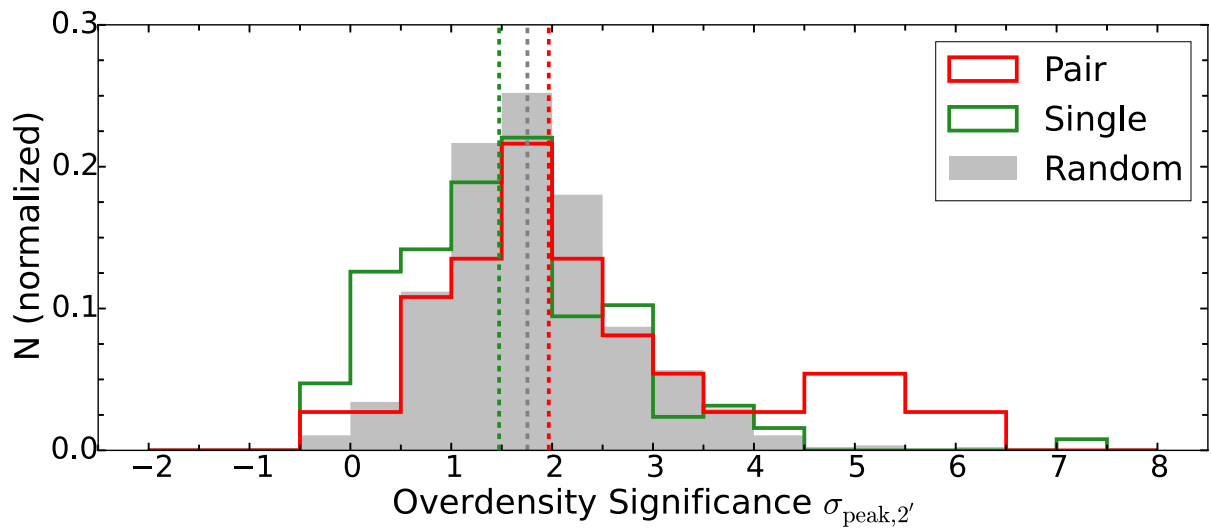} 
   \end{center}
\caption{Normalized distribution of the overdensity significance around $z<1.5$ quasar pairs (red), $z\sim1$ single quasars (green), and random galaxy fields (grey). 
The overdensity significance $\sigma_\mathrm{peak, 2'}$ is defined as the peak significance within a two-arcminutes radius from quasars or galaxies.
The pair center is used for the case of quasar pairs.
The negative significance means that all the areas inside the two-arcminutes radius have smaller galaxy densities than the average.
The median value for each distribution ($1.97$, $1.47$, and $1.75 \sigma$ for pair, single, and random, respectively) is indicated with a vertical line.}\label{fig:lowz_all}
\end{figure*}

\begin{figure*}[hp]
 \begin{center}
  \FigureFile(160mm, 160mm){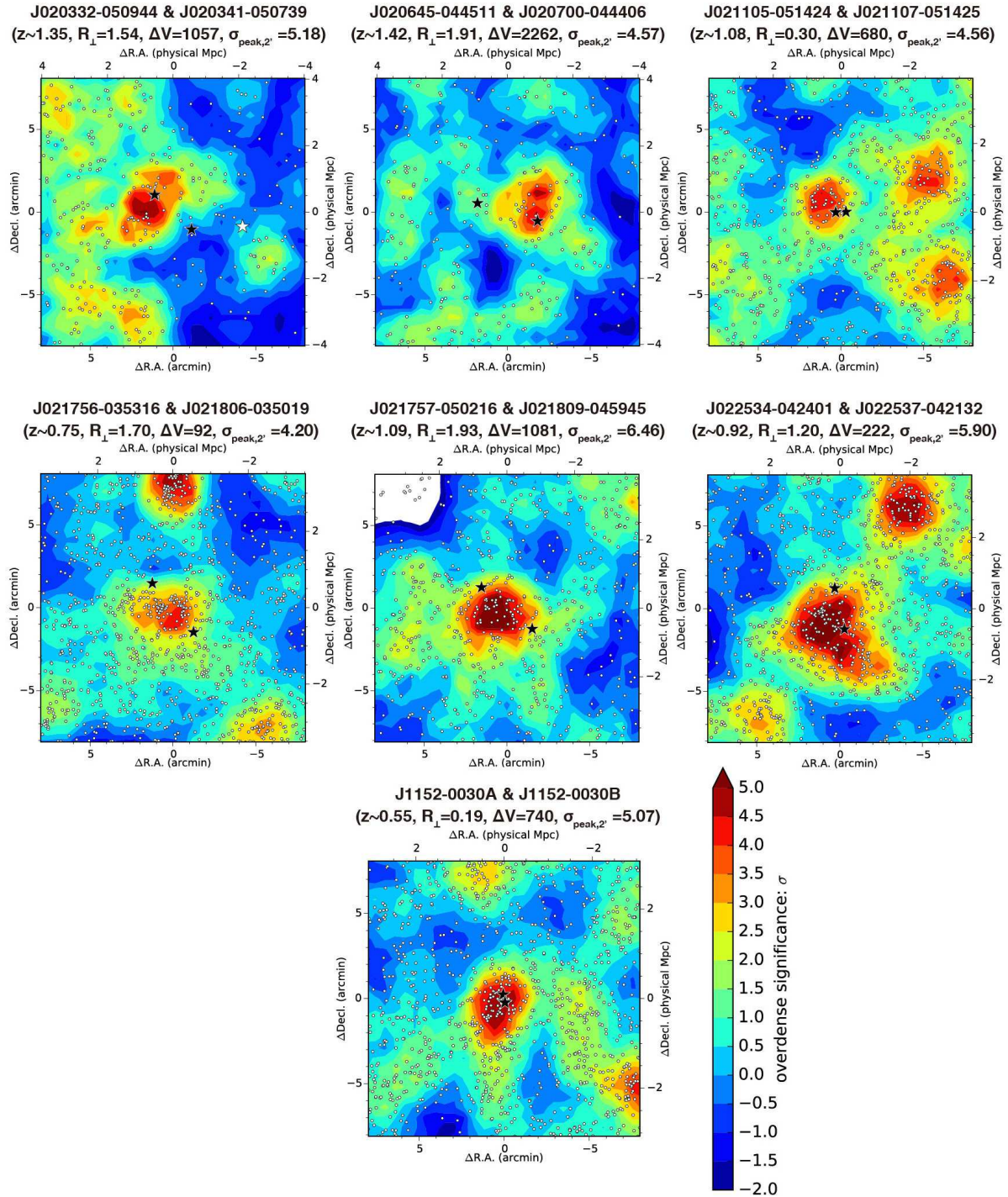} 
   \end{center}
\caption{Local significance maps for seven $z\sim 1$ pairs with $>4\sigma$ overdensity within two arcminutes. 
The symbols and contours are the same as Figure~\ref{fig:fig1}.
The first six panels show the BOSS binary fields and the bottom panel shows the one from \citet{Hennawi06}.
The white star in J020332-050944 \& J020341-050739 field shows the companion quasar (SDSSJ020320.47-050933.8) at $z=1.353$.
}\label{fig:fig_OD_R2}
\end{figure*}

\begin{figure*}[htbp]
 \begin{center}
  \FigureFile(160mm, 60mm){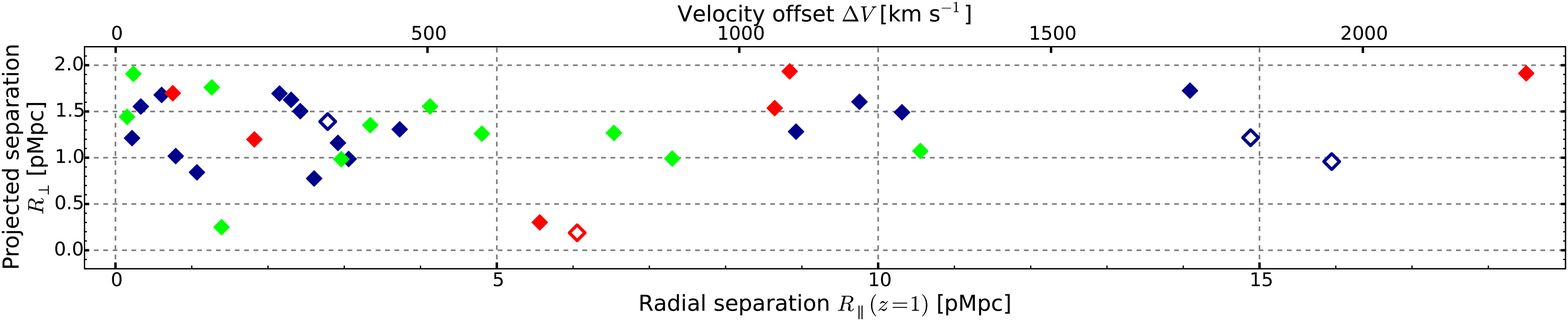} 
   \end{center}
\caption{ The distribution showing the radial separation $R_\parallel$ [pMpc] and projected separation $R_\perp$ [pMpc] of the $37$ pairs at $z\sim1$.
The radial separation $R_\parallel$ is converted from the velocity offset $\Delta V$ [km s$^{-1}$] assuming $z=1$, which is also shown for reference.
The color shows the peak significance within two arcminutes $\sigma_\mathrm{peak,2'}$ in red ($4\leq\sigma_\mathrm{peak,2'}$), green ($2\leq\sigma_\mathrm{peak,2'}<4$) and blue ($\sigma_\mathrm{peak,2'}<2$).
The filled symbols show the BOSS quasar pairs.
The open symbols show the pairs from \citet{Hennawi06}.
}\label{fig:fig_sum}
\end{figure*}

In this section, we show the result of the overdensity measurements in $z\sim1$ quasar pair fields.
Since we have an unprecedentedly large number of quasar pair samples, we are able to examine the rare pair environments with statistical approaches for the first time.
Figure~\ref{fig:lowz_all} shows the normalized distribution of the peak significance around the pairs.
Globally, the peak significance is distributed around a moderate density with the median significance of $<\sigma_\mathrm{peak,2'}>=1.97$.
This result suggests that the quasar pairs at low redshift reside in moderate environments as a whole, in contrast to the high redshift pairs.
This is consistent with the findings of the previous studies (e.g., \cite{Boris07,Sandrinelli14}) in which they find no strong evidence that the local galaxy density is highly enhanced around $z\lesssim1$ pairs.
However, it is notable that the distribution has a long tail toward high significance up to $6.46 \sigma$.
To be specific, there are seven pairs ($19\%$) with $\sigma_\mathrm{peak,2'}>4$, including SDSSJ1152-0030, the pair with the smallest projected separation ($\sigma_\mathrm{peak,2'}=5.07$), and J020332-050944 \& J020341-0500739, which has a companion ($\sigma_\mathrm{peak,2'}=5.18\sigma$).
These $>4\sigma$ pair fields were visually checked to confirm that the overdense regions are not fakes due to, for example, false detections of artificial noises around bright stars.
Their local significance maps are shown in Figure~\ref{fig:fig_OD_R2}, where it is clear that the significance is highly enhanced between or on the pair members.
The maps of the other pairs hosted in normal- or under-density regions are shown in Figure~\ref{fig:fig_A3} (Appendix~\ref{sec:lowz_A}).
We also show the peak significance in the $R_\parallel-R_\perp$ plane in Figure~\ref{fig:fig_sum}, 
where $R_\parallel$ is the line-of-sight separation directly converted from the velocity difference $\Delta V$.
The significance is divided into three bins: $4<\sigma_\mathrm{peak,2'}$, $2< \sigma_\mathrm{peak,2'}\leq 4$, and $\sigma_\mathrm{peak,2'}\leq2$ with filled symbols showing the BOSS pairs and open symbols showing the \citet{Hennawi06} pairs.
The number of pairs in each bin and the number of sub pMpc-scale pairs are summarized in Table~\ref{tab:sum}.

\begin{table}[htb]
  \tbl{Summary of $z\sim1$ pair environments}{%
  \begin{tabular}{ccccc}
      \hline
      &$N_{4\leq\sigma_\mathrm{peak, 2'}}$&$N_{2\leq\sigma_\mathrm{peak, 2'}<4}$&$N_{\sigma_\mathrm{peak, 2'}<2}$&$N_\mathrm{total}$\\     
      \hline\hline
      $R_\perp<1$&2&3&4&9\\  
      $R_\perp\geq1$&5&8&15&28\\  
      \hline\hline
    \end{tabular}}\label{tab:sum}
\begin{tabnote}
{\bf Notes:} The overdensity significance of the 37 $z\sim1$ pair fields are divided by the projected separation of the pair ($R_\perp<1$ and $R_\perp\geq1$ [pMpc]).
\end{tabnote}
\end{table} 
In Figure~\ref{fig:lowz_all}, we also compare the significance distribution with those of single quasars and randomly selected galaxies.
As the median significances are $1.47$ and $1.75 \sigma$, respectively, the overall significant distributions are only different by $<1\sigma$ level from the pairs.
However, a major difference is at the high density outskirt, where the significance distributions of single quasars and galaxies decline smoothly
with the fraction of $>4\sigma$ significance regions $2.4$\% and $2.0$\%, respectively.
Taking advantage of the large sample size, we perform two-sample tests of goodness-of-fit to compare the three distributions.
In all tests, the null hypothesis is that two non-parametric distributions are from the same underlying distribution.
The significance threshold is set at $0.05$.
Three combinations of the pairs (``P"), single quasars (``S"), and random fields (``R") distributions are tested as we summarize in Table~\ref{tab:stats}.
First, the Kolmogorov-Smirnoff (KS) test \citep{KS} shows that we cannot reject the possibility that any of the three samples come from the same distribution, implicating that there seem to be no significant levels of overdensity enhancement in pair and single quasar fields in a global view.
On the other hand, the Anderson-Darling (AD) test \citep{AD}, which is more sensitive to the tail of the distribution, statistically supports that the pair environments are more likely to be overdense.
Moreover, comparing the two quasar groups (``P-S" in Table~\ref{tab:stats}), we find that quasar pairs favor cluster environments.
Intriguingly, it is also evident that the single quasar fields are likely underdense, compared with the galaxy fields. 
Since the two pairs at $z=3.3$ and $3.6$ have physical separations comparable to the $z\sim1$ pairs,
our result suggests that $<2$ pMpc-scale quasar pairs are good tracers of massive clusters both at $z>3$ and $z\sim1$,
yet the probability of finding clusters seems smaller at low redshift.
\begin{table}[ht]
  \tbl{Two-sample KS and AD Test}{%
  \begin{tabular}{cccccc}
      \hline
& \multicolumn{2}{c}{KS} && \multicolumn{2}{c}{AD}\\
\cline{2-3}\cline{5-6}
      &$D$&$p$&&$A^2$&$p$\\     
      \hline\hline
      P-R&0.30&0.28&&5.1&0.0033\\  
      S-R&0.20&0.77&&9.8&0.0001\\  
      P-S&0.30&0.28&&6.3&0.0014\\  
      \hline
      P$_{z1.0}$ - P$_{z1.5}$&0.20&0.77&&-1.0&1.0\\  
      \hline\hline
    \end{tabular}}\label{tab:stats}
\begin{tabnote}
{\bf Notes:} ``P" represents the quasar pairs, while ``S" and ``R" represent the single quasars at $0.9<z<1.1$ and the random sample as described in Section~\ref{sec:lowz_random}. For example, the ``P-S" stands for the comparison of the pair and single quasars.
The ``P$_\mathrm{z1.0}$-P$_\mathrm{z1.5}$" is the comparison of the pairs divided into two groups: $z<1.0$ and $1.0\leq z<1.5$.
The test statistics are shown in $D$ and $A^2$ with corresponding $p$-values.
\end{tabnote}
\end{table}

\subsection{Significance dependence on redshift}\label{sec:lowz_pair_comp_z}
\begin{figure*}[htbp]
 \begin{center}
  \FigureFile(160mm, 50mm){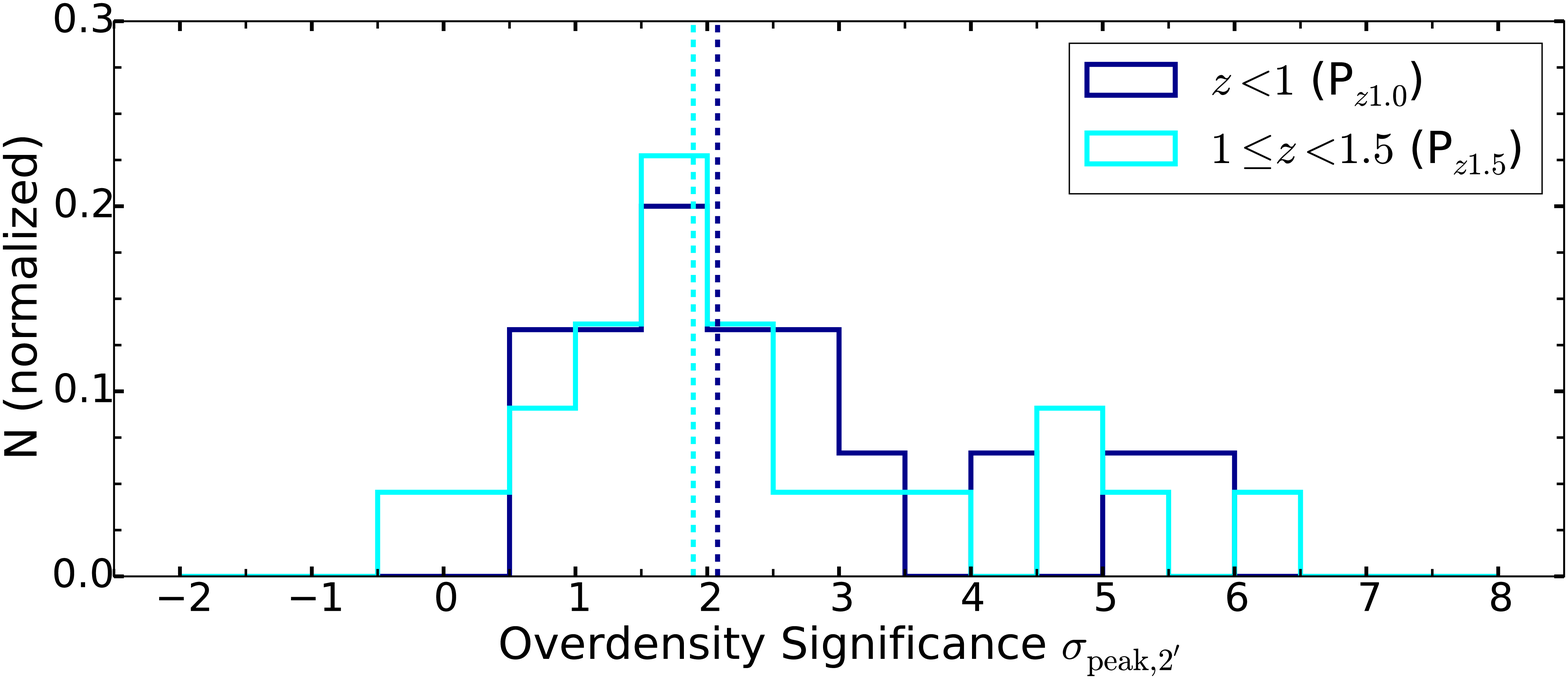} 
   \end{center}
\caption{Normalized significance distribution of the $z\sim1$ pairs divided into $z<1.0$ pairs (``P$_{z1.0}$", blue) and $1.0\leq z<1.5$ pairs (``P$_{z1.5}$", cyan).
There are $15$ pairs in P$_{z1.0}$ group and $22$ pairs in P$_{z1.5}$ group.
The grey histogram shows the random sample which is the same as in Figure~\ref{fig:lowz_all}.
}\label{fig:fig5}
\end{figure*}

 Here, we examine the redshift dependence, albeit narrow range, of the peak significance of the pair environments.
The pair sample is divided into i) $z<1.0$ group and ii) $1.0\leq z<1.5$ group.
There are $15$ pair fields at $z<1$ with median significance $2.08\sigma$
and $22$ pair fields at $1\leq z<1.5$ with median significance $1.90\sigma$.
Figure~\ref{fig:fig5} compares the normalized significance distributions of the two pair groups. 
After applying the two-sample tests, we find that there is no significant redshift dependence of the peak significance (Table~\ref{tab:stats}).
Note that this result supports our initial assumption that the pair environments do not significantly change at $0.3<z<1.5$.

\section{Discussion}\label{sec:discussion}
\subsection{Enhancement of overdensity around $z>3$ quasar pairs}\label{sec:discussion_highz}
In this study, we derive an implication that 
the rare occurrence of $<2$ pMpc-scale quasar pairs is related to galaxy overdensity regions at $z>3$ and also, with a statistical evidence at $z\sim1$.
At the high redshift, if the effective projection size of a $z\sim4$ proto-cluster is defined as $1.8$ arcminute ($0.75$ pMpc, \cite{Chiang13}) radius,
the total surface area of the $179$ HSC proto-clusters is $0.51$ deg$^2$, only $0.4\%$ of the entire S16A field.
Therefore, when one assumes a uniform surface density of quasar pairs, 
the chance that two randomly selected positions in $121$ deg$^2$ are both in proto-cluster fields is only $2\times 10^{-5}$, while we find two proto-clusters out of the two pairs.

At this stage, we again look at the work of \citet{Uchiyama17}.
Using the same HSC $g$-dropout selection and parent quasar catalog, they suggest that the majority of $151$ quasars at $z\sim3.8$ likely reside in moderate-density environments.
As also mentioned in Section~\ref{sec:results_z34} in this paper, 
there are only six BOSS quasars located within three arcminutes ($\sim1.25$ pMpc) from the density peaks of the HSC proto-clusters, 
and intriguingly, three of them are the QSOP1 and QSOP2 pair members (Figure~\ref{fig:fig1}).
While several studies argue high-redshift quasar environments with small-field observations, for example using the {\it Hubble Space Telescope}, the area used for the overdensity measurements in our HSC-SSP studies is sufficient to identify proto-clusters and to see their whole structures thanks to the enormously wide coverage.
Thus, it is assured that, unlike the average environments of isolated quasars at $z\sim3.8$, the two quasar pair fields showing large excess ($>5\sigma$) of overdensity significance are exceptionally rare and rich as quasar environments at this epoch.
In terms of the triggering mechanism, our result suggests that, even though the vast majority of luminous quasars may be triggered via the secular process which is independent of the galaxy density, quasars turning on by the heavy interaction of massive galaxies do exist in massive haloes, and such quasar activity is perhaps so synchronous especially at high redshift that we observe more than one quasar in proto-clusters.
This result supports our initial anticipation that pairs of luminous quasars 
are better tracers of proto-clusters than single quasars.
Meanwhile, given that \citet{Fukugita04} find no overdensity around a quasar pair at $z=4.25$, 
 while their imaging area is small ($5.8$ arcminutes$^2$),
a bigger pair sample at $z>3$ is needed to derive more conclusive evidence.
Spectroscopic or narrow-band imaging follow-up observation of the two pair fields is required to address
i) the identification of member galaxies, ii) the presence of faint quasars as the one found in the QSOP2 field, and iii) star-formation activity of the member galaxies.

It is notable that the two pair fields are not the richest among the HSC proto-clusters.
\citet{Toshikawa17} find up to $9\sigma$ overdensity regions, while the pair fields have $4.8$ and $4.0\sigma$ in their measurements.
The interpretation of this fact is not straightforward. 
The AGN feedback suppression of the quasar activity in the most massive haloes \citep{Fanidakis13, Orsi15} could explain the absence of luminous quasars in the high density regions, though it cannot explain why the QSOP1 and QSOP2 pair members can emerge in proto-clusters.
A certain number of quasars in proto-clusters could be missed in the incomplete selection of the BOSS catalog especially at $z>3.5$, but
it is unlikely that the members of the two quasar pairs pass the BOSS color selection but the majority of other quasars in other $>100$ proto-clusters do not.
As an alternative explanation, AGN in the most massive haloes just might not turn on in the more massive proto-clusters, if one takes into account the short AGN duty cycle of typically $10^7$ years.
In this context, another possibility would be that the quasar activity is triggered in the massive haloes but they are not in the type-I quasar phase; thus they are not optically bright.
Since we are just starting the proto-cluster search in the first $\sim100$ deg$^2$, it would be of great interest to study the relation between the most overdense regions and luminous quasars with a larger sample size as the HSC-SSP survey coverage is enlarged.
Furthermore, far-infrared or sub-mm studies of the HSC proto-clusters are necessary to probe obscured galaxies residing in proto-clusters.

We note that two sub pMpc-scale quasar pairs in \citet{Hennawi10}, namely SDSSJ1054+0215 at $z\sim3.98$ and SDSSJ1118+0202 at $z\sim3.94$ are at our redshift range and within the entire HSC-Wide survey area, while these fields have not been covered yet.
Also, CFHTLSJ0221-0342, another small-scale pair at $z=5$ reported in \citet{McGreer15} is within the W-XMMLSS field covered with the HSC-Deep layer, but the current depth is almost the same as the Wide.
We will investigate these small-scale pair environments at $z>3$ after their fields are covered or the imaging gets deeper, to compare with larger separation pairs like the two in this paper as well as the other pairs which will be covered in the future HSC-SSP data release, and also with the $z\sim1$ counterparts.

Finally, the environment around fainter $z>3$ quasars is another important topic.
While low-luminosity quasars are more easily triggered in normal environments as would also be the case for quasar pairs,
such measurements will provide a clue to understand how the host haloes and surrounding galaxy overdensity affect the triggering mechanism of low-luminosity quasars.
In the HSC-SSP, \citet{Akiyama17} compile a $>1000$ sample of $z\sim4$ quasars down to $i=24$ to derive the accurate shape of the quasar luminosity function at the faint end.
\citet{He17} discuss the clustering of their quasars to show that their host halo masses are moderate, with the order of $M_\mathrm{halo}\sim10^{12}M_\odot$.
To asses their measurements from an environment point of view, it is highly necessary to investigate the galaxy overdensity of their low-luminosity quasars.

\subsection{Redshift dependence of quasar pair environments}\label{sec:discussion_lowz}
Different environments between single quasars and pairs are also found at $z\sim1$, as we show in Section~\ref{sec:lowz_pair_res} that a significant fraction of quasar pair fields is at high density regions, which is statistically different from those of single quasars and randomly selected galaxies.
However, a big difference from the $z>3$ pair environments is that the significance distribution is globally the same among the three environments, meaning that quasars are generally common in any environments at $z\sim1$.
Since the brightness of most $z\sim1$ quasars is in the range of $19<i<21$ and this is the same for the pairs in $>4\sigma$ regions, the brightness dependence of the pair environments does not have a major effect on the result.

The trend that not all $z\sim1$ quasar pairs reside in massive environments as well as single quasars is consistent with previous studies.
For example, \citet{Farina11} show that, although the pair selection is severer than this study (i.e., $R_\perp<500$ kpc and $\Delta V<500$ km s$^{-1}$), in only one out of six pair fields at $z<0.8$ shows significant overdensity using galaxies as bright as the ones in this study.
\citet{Boris07} adopt a loose pair selection at $z\sim1$ ($\Delta \theta<300$ arcseconds) comparable to this study.
They conduct a deep optical imaging observation with Gemini/GMOS down to $1.5\sigma$ limit $i'=26.4$ in four pair fields, showing that three pairs are associated with cluster environments but the other one is in an isolated field.
\citet{Sandrinelli14} stack the radial profile of galaxies in $14$ pair fields at $z<0.85$ ($R_\perp<600$ kpc) and suggest that there is no clear enhancement of overdensity compared with single quasars.
We note that the definition of overdensity and its significance are different among this and previous studies; therefore the probability of finding overdense regions around quasar pairs cannot be compared with this study.
These results can be discussed under the downsizing evolution of the quasar activity.
At high redshifts, quasar activity is most efficient in the most massive SMBHs residing in dense environments, for which plenty of cold gas is available for the mass growth to make gigantic SMBHs ($M_\mathrm{BH}>10^9M_\odot$) such as those found at $z>6$ (e.g., \cite{Mortlock11, Wu15}).
At lower redshifts, the galaxy interaction gets common in normal environments following the fast growth of SMBHs in proto-clusters.
The luminosity where the quasar activity is most active shifts to a less luminous range as quasars are powered by already-matured black holes or less-massive black holes in their late growth.
The observational evidence of such anti-hierarchical evolution is given by, for instance, \citet{Ueda14} in their X-ray AGN luminosity function over $0<z<5$.
In this respect, our result confirms the down-sizing trend of the SMBH growth by showing from an environmental point of view that the majority of low-redshift quasars, even for quasar pairs, turn on in moderate environments, in contrast to the $z>3$ quasar pairs.

Furthermore, our measurement of $>30$ $z\sim1$ pair environments shows statistical evidence that, not all but $\sim20\%$ of the quasar pairs does reside in cluster fields and this is distinguishable from the isolated quasar eivironments and random fields.
This feature is found thanks to our large sample size, while previous studies are limited to $\sim10$ pair fields.
The high density tail in the significance distribution of the pairs suggests that the quasar activity is still ongoing in massive haloes and actually so active at $z\sim 1$ that more than one quasar are triggered at the same time.
There should be isolated quasars in other massive environments, but they would be not visible since they are hidden by larger number of quasars in moderate environments.
The implication of this trend is that there are still remnants and further accretion of cold-gas in cluster environments even after the major epoch of star-forming and SMBH feeding, and the frequent galaxy interaction can ignite more than one quasar simultaneously.

There are nine quasar pairs with less than $1$ pMpc projected separation including two from \citet{Hennawi06}.
From the nine sub pMpc-scale pairs, we find that two pairs, J021105-051424 \& J021107-051425 at $z\sim1.08$ ($R_\perp=0.30$ pMpc, $\Delta V=680$ km s$^{-1}$) and J1152-0030A \& B at $z\sim0.55$ ($R_\perp=0.19$ pMpc, $\Delta V=740$ km s$^{-1}$) are embedded in high density environments ($\sigma_\mathrm{peak,2'}=4.56\sigma$ and $5.07\sigma$, respectively).
On the other hand, the significance of the other seven pairs is moderate or small ($\sigma_\mathrm{peak,2'}=0.4-3.0$) like the larger-separation pairs.
Therefore, while several studies on such small-scale clustering of quasars suggest that the clustering signal of projected correlation function is enhanced from the extrapolation from Mpc scale due to intense interaction of galaxies in massive haloes \citep{Eftekharzadeh17, Hennawi06},
our result implies that such small-scale quasar pairs are hosted not only in cluster fields but also in general fields at $z\sim1$.

\section{Summary}\label{sec:summary}
In this paper, we investigate the galaxy overdensity of quasar pair environments at high ($z>3$) and low ($z\sim1$) redshift, using the optical imaging catalog of the HSC-SSP survey (DR S16A) covering effectively $121$ deg$^2$ with the $i$-band $5\sigma$-depth of $\sim26.4$. 
The quasar pairs are primarily extracted from the SDSS DR12Q catalog with additional samples from the literature at $z\sim1$.
We use the photometric catalog of the HSC-SSP to select surrounding galaxies based on our $g$-dropout selection at $z>3$ and photometric redshift selection at $z\sim1$.
The galaxy overdensity measurement around quasar pairs is based on the local significance of galaxies within a $1.8$ arcminutes aperture at each square grid over a $2\times2$ deg$^2$ field separated by $0.6$ arcminute.
Our main results are summarized as follows:

\begin{enumerate}
\item We find that two quasar pairs at $z=3.3$ and $3.6$ are associated with $\sigma_\mathrm{peak}>5\sigma$ overdense regions.
Their projection separations are $R_\perp=1.75$ and $1.04$ pMpc, and their velocity offsets are $\Delta V=692$ and $1448$ km s$^{-1}$, respectively.
The number counts within five arcminutes around the two pairs are more than twice as high as field environments in all magnitude bins down to $i=26$.
Specifically, QSOP1 field has two brightest dropouts ($21.2\leq i\leq21.4$), which is six times higher density than the general fields.
Since no apparent trend has been found to indicate that quasars at $z\sim3.8$ are related to massive environments, pairs of luminous quasars are likely more efficient tracers of proto-clusters than isolated quasars,
although a bigger pair sample is needed to discuss the probability of tracing the high density regions and its redshift dependence.
Our result implies that the two quasar pairs are likely triggered via galaxy major merger, while the vast majority of isolated quasars are triggered via other processes such as bar and disk instabilities.
\item The overdensity significance of the $z>3$ pair environments is not the highest among $179$ HSC proto-clusters, which may imply that luminous quasars cannot emerge in the most massive haloes.
However, this is not clear since we may miss quasars in the largest proto-clusters due to the incomplete selection of $z>3.5$ quasars, and quasars may actually exist in the richest environments but they are not optically bright due to obscuration or their turn-off phase in the duty cycle.
\item We extend our study down to $z\sim1$ using the HSC-SSP photometric redshifts. 
We select $33$ pairs from the BOSS DR12Q and also four previously known small-scale pairs from \citet{Hennawi06}.
While the distribution of peak significance within two arcminutes from the pair center is globally not different from those of isolated quasars and randomly selected galaxies at the same redshift range, a significant difference is found at the high density tail thanks to our large sample size.
We find that $19\%$ of the $z\sim1$ pairs are within massive ($>4\sigma$) environments and statistically confirm that this is unique in pair environments.
Our result suggests that more than one quasar can ignite simultaneously in massive haloes even after the major epoch of the AGN activity, and quasar pairs are still good tracers of rich environments at $z\sim1$, while the chance is lower than at $z>3$.
We detect no redshift dependence of the significance between $z<1$ pairs and $1\leq z<1.5$ pairs.
\item Among nine small-scale pairs with $R_\perp<1$ pMpc,
two of them reside in $>4\sigma$ fields including the pair with the smallest projected separation, SDSSJ1152-0030A \& B ($z\sim0.55$, $R_\perp=0.19$ pMpc).
The other fields are moderate- or under-density regions, suggesting that sub pMpc-scale pairs could be embedded in general fields at low-redshift like isolated quasars.
\end{enumerate}

\begin{ack}

We thank the anonymous referee for helpful comments on the manuscript.
MO would like to express gratitude to B. Venemans and E. P. Farina for fruitful discussions 
at the initiation of this work.
MO acknowledges Roderik Overzier, Toshihiro Kawaguchi and Satoshi Kikuta for the suggestions they offered for the manuscript.

This work is based on data collected at the Subaru Telescope and retrieved from the HSC data archive system, which is operated by Subaru Telescope and Astronomy Data Center at National Astronomical Observatory of Japan.

The Hyper Suprime-Cam (HSC) collaboration includes the astronomical
communities of Japan and Taiwan, and Princeton University.  The HSC
instrumentation and software were developed by the National
Astronomical Observatory of Japan (NAOJ), the Kavli Institute for the
Physics and Mathematics of the Universe (Kavli IPMU), the University
of Tokyo, the High Energy Accelerator Research Organization (KEK), the
Academia Sinica Institute for Astronomy and Astrophysics in Taiwan
(ASIAA), and Princeton University.  Funding was contributed by the FIRST 
program from Japanese Cabinet Office, the Ministry of Education, Culture, 
Sports, Science and Technology (MEXT), the Japan Society for the 
Promotion of Science (JSPS),  Japan Science and Technology Agency 
(JST),  the Toray Science  Foundation, NAOJ, Kavli IPMU, KEK, ASIAA,  
and Princeton University.

This paper makes use of software developed for the Large Synoptic Survey Telescope. We thank the LSST Project for making their code available as free software at  {\tt http://dm.lsst.org}\\
The Pan-STARRS1 Surveys (PS1) have been made possible through contributions of the Institute for Astronomy, the University of Hawaii, the Pan-STARRS Project Office, the Max-Planck Society and its participating institutes, the Max Planck Institute for Astronomy, Heidelberg and the Max Planck Institute for Extraterrestrial Physics, Garching, The Johns Hopkins University, Durham University, the University of Edinburgh, Queen's University Belfast,  the Harvard-Smithsonian Center for Astrophysics, the Las Cumbres Observatory Global Telescope Network Incorporated, the National Central University of Taiwan, the Space Telescope Science Institute, the National Aeronautics and Space Administration under Grant No. NNX08AR22G issued through the Planetary Science Division of the NASA Science Mission Directorate, the National Science Foundation under Grant No. AST-1238877, the University of Maryland, and Eotvos Lorand University (ELTE) and the Los Alamos National Laboratory.

A part of this work is financially supported by the overseas internship fund of Department of Astronomical Science, SOKENDAI and JSPS KAKENHI Grant Number 15J02115.
NK acknowledges supports from the JSPS KAKENHI Grant Number 15H03645.
\end{ack}

\appendix 
\section{Low density environments around $z\sim1$ quasar pairs}\label{sec:lowz_A}
In Section~\ref{sec:lowz_pair_res}, we show that seven out of $37$  $z\sim1$ quasar pairs are embedded in $>4\sigma$ overdenisty regions as shown in Figure~\ref{fig:fig_OD_R2}.
Here, we show the other $29$ pair fields showing lower ($\sigma_\mathrm{peak,2'}<4$) overdensity significance for reference in Figure~\ref{fig:fig_A3}, in which their redshift, pair separation ($R_\perp$ and $\Delta V$), and peak significance $\sigma_\mathrm{peak,2'}$ are given.
Note that J020320-050933 \& J020332-050944, the companion pair of J020332-050944 \& J020341-050739 at $z\sim1.35$ is shown in the top-left panel of Figure~\ref{fig:fig_OD_R2}.

\begin{figure*}[h]
 \begin{center}
  \FigureFile(160mm, 160mm){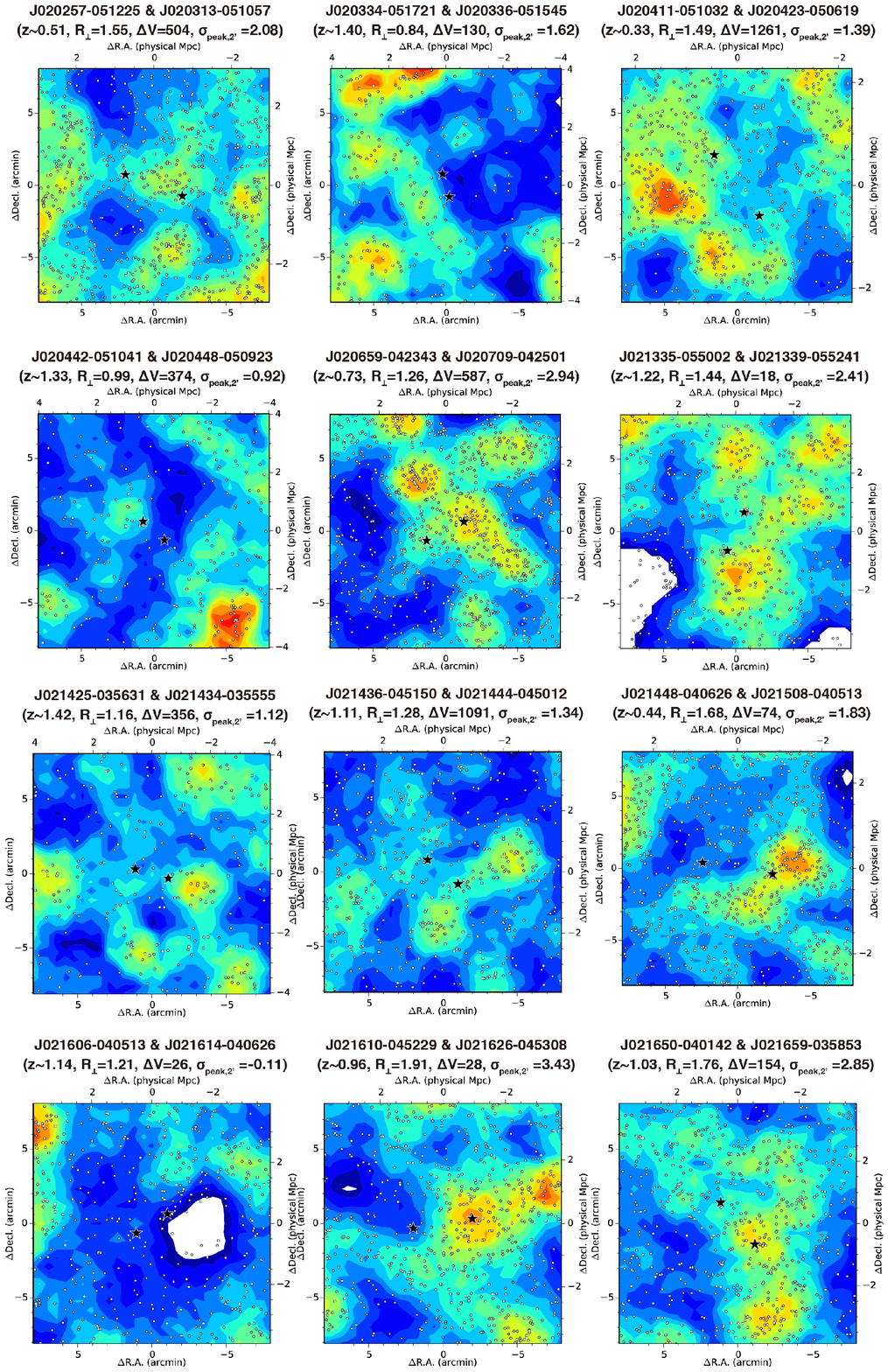} 
   \end{center}
\end{figure*}
\begin{figure*}[h]
 \begin{center}
  \FigureFile(160mm, 160mm){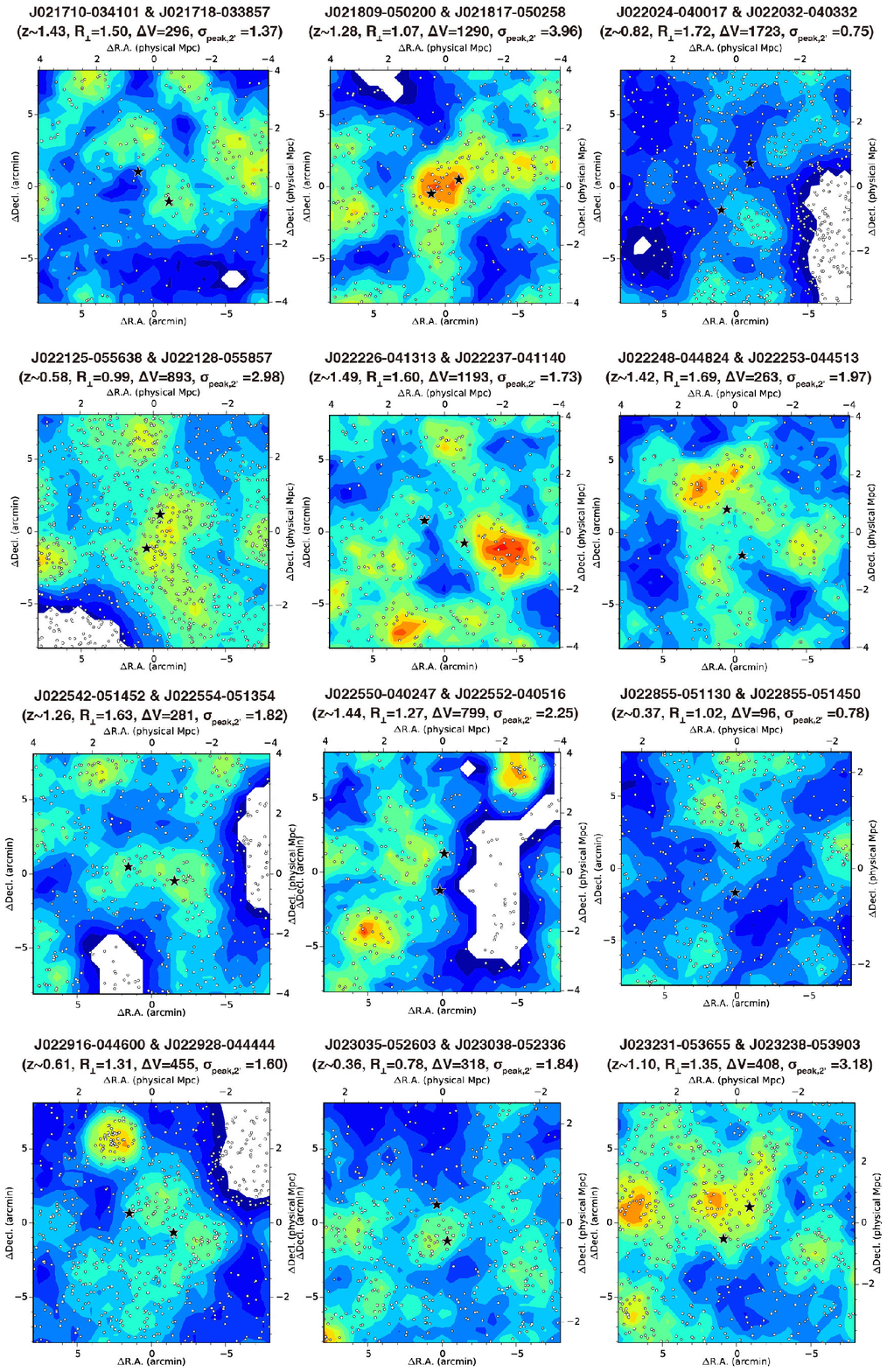} 
   \end{center}
\end{figure*}
\begin{figure*}[htb]
 \begin{center}
  \FigureFile(160mm, 160mm){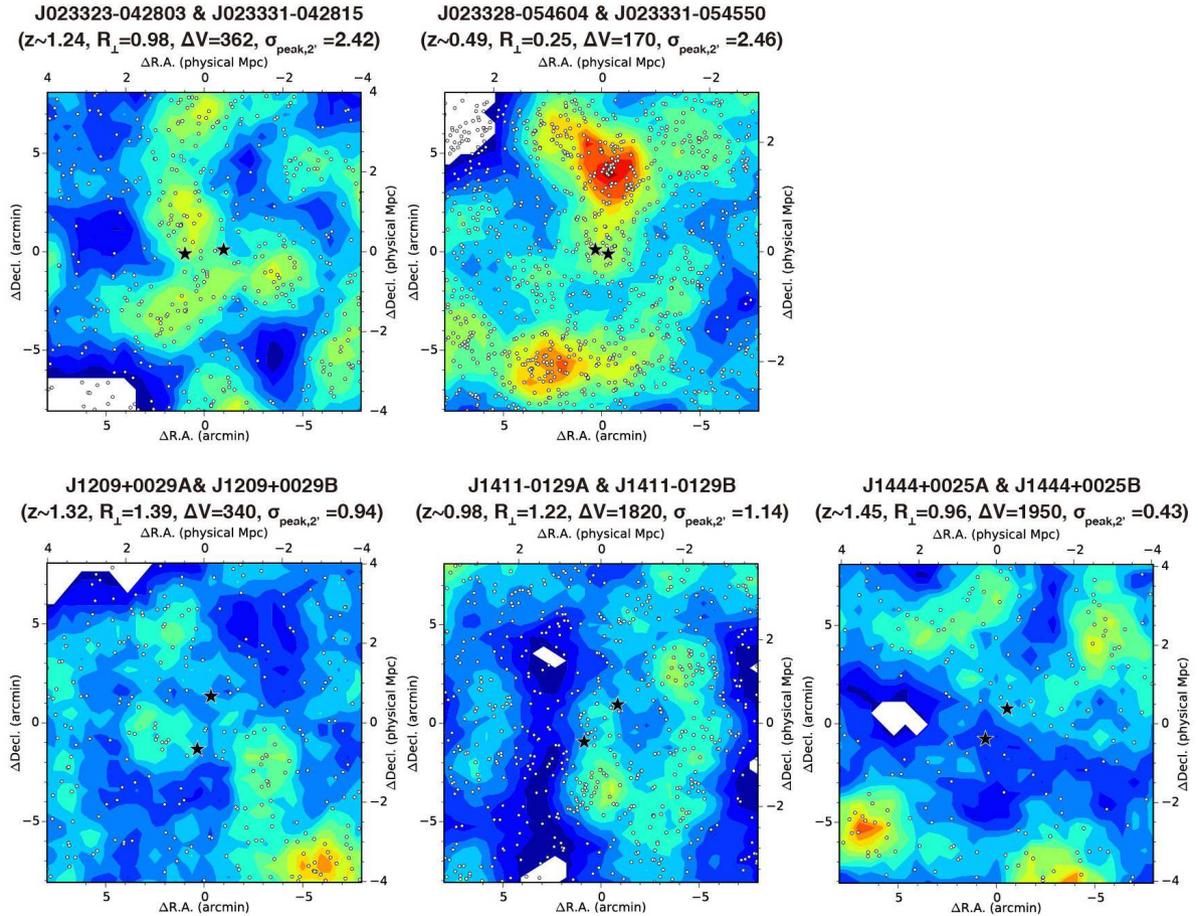} 
   \end{center}
\caption{
Local significance maps of the $30$ pair fields showing $\sigma_\mathrm{peak,2'}<4\sigma$.
The symbols and contours are the same as Figure~\ref{fig:fig_OD_R2}.
The first $27$ panels show the BOSS pairs (see Table~\ref{tab:tab_lowz}), and the last three are the \citet{Hennawi06} pairs (see Table~\ref{tab:tab_lowz_b}).
Note that one field J020320-050933 \& J020332-050944-050944 is shown in the panel in Figure~\ref{fig:fig_OD_R2} (the first panel at the left-top).
}
\label{fig:fig_A3}
\end{figure*}
%
%



\end{document}